\documentstyle[12pt]{article}
\newcommand{\pr}{\paragraph{}}
\newcommand{\be}{\begin{equation}}
\newcommand{\ee}{\end{equation}}
\newcommand{\bea}{\begin{eqnarray}}
\newcommand{\nn}{\nonumber}
\newcommand{\eea}{\end{eqnarray}}
\newcommand{\nd}[1]{/\hspace{-0.6em} #1}
\newcommand{\nk}{\noindent}
\begin{document}

\begin{titlepage}
\begin{flushright}
CERN-TH.6897/93 \\
CTP-TAMU-30/93 \\
ACT-10/93 \\
\end{flushright}
\begin{centering}
\vspace{.05in}
{\large {\bf A String Derivation of the  $\nd{S}$
Matrix}} \\
\vspace{.1in}
{\bf John Ellis$^a$}, {\bf N.E. Mavromatos$^{a,b}$}
and {\bf D.V. Nanopoulos$^{a,c}$}\\
\vspace{.03in}
\vspace{.1in}
{\bf Abstract} \\
\vspace{.05in}
\end{centering}
{\small
We show that, in
string theory, as a result
of the $W_{\infty}$-symmetries
that preserve quantum coherence
in the {\it full} string theory
by coupling different mass levels,
transitions between initial- and final-state
density matrices for the effective
light-particle theory
involve non-Hamiltonian
terms $\nd{\delta H}$
in their time evolution, and are described
by a $\nd{S}$ matrix that is not factorizable
as a product of field-theoretical
$S$ and $S^\dagger$ matrices. We exhibit
non-trivial
string
contributions to $\nd{\delta H}$ and
the
$\nd{S}$ matrix associated with topological fluctuations
related to the
coset model that describes an s-wave black hole. These include
monopole-antimonopole configurations on the world-sheet
that correspond to black hole creation and
annihilation, and
instantons that
represent back-reaction via
quantum jumps between black holes of different mass,
both of which
make the string supercritical. The resulting Liouville mode is
interpreted as the time variable, and the arrow of time is
associated with black hole decay. Since conformal invariance is
broken in the non-critical string theory, monopole and antimonopole, or
instanton and
anti-instanton, are not separable, and the valley trajectories
between them contribute
to the $\nd{S}$ matrix. Ultraviolet
divergences in the valley actions
contribute non-Hamiltonian terms $\nd{\delta H}$
to the time evolution of the density
matrix, causing it to become mixed and suppressing
off-diagonal interference terms
in
the configuration-space representation of the
final-state density matrix. This can be understood as a consequence
of the fact that the arrow of time causes world-sheets to
contract in target space.}

\paragraph{}
\par
\vspace{0.05in}
\begin{flushleft}
CERN-TH.6897/93 \\
CTP-TAMU-30/93 \\
ACT-10/93 \\
May 1993 \\
\end{flushleft}
\vspace{.01in}
--------------------------------------------- \\
$^{a}$ Theory Division, CERN, CH-1211, Geneva 23, Switzerland  \\
$^{b}$ Laboratoire de Physique Th\`eorique
ENSLAPP (URA 14-36 du CNRS, associe\`e \`a l' E.N.S
de Lyon, et au LAPP (IN2P3-CNRS) d'Annecy-le-Vieux),
Chemin de Bellevue, BP 110, F-74941 Annecy-le-Vieux
Cedex, France. \\
$^{c}$ Center for
Theoretical Physics, Dept. of Physics,
Texas A \& M University, College Station, TX 77843-4242, USA
and, Astroparticle Physics Group, Houston
Advanced Research Center (HARC),The Woodlands, TX 77381, USA.

\end{titlepage}
\newpage
\section{Introduction and Summary}
\pr
The two greatest revolutions in physics in the first half of the
twentieth century were those leading to quantum mechanics and
general relativity. The greatest piece of unfinished business for
physics in the second half of the century is their reconciliation
in a consistent and complete quantum theory of gravity. Such a
synthesis is likely to require that either or both of quantum
mechanics and general relativity be modified in an essential
way. It has been argued
that any consistent quantum theory of
gravity must indeed modify traditional quantum field theory
\cite{hawk} and quantum
mechanics\cite{ehns} in such
a way as to accommodate the passage from pure to mixed states that
seems to be unavoidable when information is lost across an event
horizon \cite{bek}, for example that surrounding a black hole, whether
macroscopic or microscopic.
\pr

The
original argument was presented in the context of
conventional point-like field theory \cite{hawk}.
However, there are other
reasons, such as the appearance of uncontrollable infinities in
perturbation theory, to think that any point-like quantum
theory is doomed. Indeed, the only candidate we have for a
consistent quantum theory of gravity is string theory, in which
point-like particles are replaced by extended objects. Thus
general relativity is modified in an essential way by this candidate
for a complete quantum theory of gravity. The next question
within the string framework
is
whether conventional quantum mechanics is also modified in an
essential way \cite{emnqm}.
\pr

It has been
suggested \cite{hawk}
that the conventional $S$-matrix description of
scattering in quantum field theory would need to be abandoned in
quantum gravity, that one should
work in a density matrix
formalism, and consider the superscattering
$\nd{S}$ matrix that relates the
outgoing density matrix to the incoming one,
\be
\rho _{out} = \nd{S} \rho _{in}
\label{dollar}
\ee
In conventional
quantum field theory,
the $\nd{S}$ matrix is factorizable as the product of an $S$ and
an $S^\dagger$ matrix: $\nd{S}$ = $S$ $S^\dagger$, but
some model calculations \cite{hawk1} indicated
that the $\nd{S}$ matrix
would not factorize in a true quantum theory of gravity. This
argument has been questioned in the literature \cite{gross}:
in this paper we
demonstrate the correctness of the
conclusion in the
context of string theory.
\pr
Some time ago, two of us (J.E. and D.V.N.), together with J.S. Hagelin
and M. Srednicki, pointed out\cite{ehns}
that if asymptotic scattering was
described by a non-factorizable $\nd{S}$ matrix as in equation (1),
there should be a corresponding modification of the quantum
Liouville
equation which describes the continuous
evolution of the density matrix when averaged over time scales
long compared with the Planck time:
\be
\partial _t \rho \equiv {\dot \rho} =
i[ \rho, H] +     \delta \nd{H}\rho
\label{modliouv}
\ee
This form of equation is characteristic of an open
quantum-mechanical system, and expresses the intuition that
observable particles are coupled to degrees of freedom
that are associated with microscopic space-time singularities
and unobservable in conventional laboratory experiments.
Subsequently, J.E. and D.V.N. together with S. Mohanty
showed \cite{emohn}
that
the modification (\ref{modliouv}) of the quantum Liouville equation
caused off-diagonal interference terms in the density matrix
to collapse over time, and that this collapse could be
very rapid for large objects\cite{emohn2}
 such as Schroedinger's cat.
\pr

During the last couple of years we have been
investigating\cite{emn1,emn4d,emnloop,emnsel,emndua,emncpt}
whether modifications of quantum mechanics of the forms
(\ref{dollar}) or (\ref{modliouv}) appear
in string theory, using the s-wave stringy black
hole as a theoretical laboratory\footnote{As we discuss
elsewhere\cite{emnvs}, we do not consider
a two-dimensional dilaton gravity field theory
to be an adequate model for studying black hole physics
in string theory.}. We have found that quantum
mechanics is maintained in a fixed black hole background,
because of an infinite-dimensional $W$-algebra which inter-relates
string states with different masses\cite{emn1}. Its
Cartan subalgebra provides an infinite set of conserved, non-local
gauge charges that label all the black hole states and enable
the information apparently lost across the event horizon to be
encoded and retained\cite{emn1}. Thus
there is a well-defined $S$ matrix
for the scattering of light particles off a black hole, black
hole decay is a quantum-mechanical effect of higher
genera\cite{emnloop},
not a thermodynamic effect, and we have proposed gedanken
experiments capable in principle of measuring all the $W$
quantum numbers of the black hole\cite{emnsel}.
However, this does not mean that quantum mechanics
applies unmodified to conventional laboratory experiments, in
which these quantum numbers are not measurable
at the microscopic level.
\pr
It is necessary \cite{emnqm}
to extend the discussion of fixed space-time
backgrounds to include quantum fluctuations, namely
space-time foam, which includes microscopic black holes that
are described by a statistical gas of
monopoles and vortices on the world-sheet\cite{emndua}.
A laboratory
experiment conducted with light particles does not measure
the massive degrees of freedom associated with these microscopic
black holes. However, light and heavy states are coupled via
the $W$-algebra\cite{emn1}, which
expresses the back-reaction of particles
on the background metric. Laboratory measurements therefore
truncate string theory in a manner incompatible with the
coherence-preserving $W$ symmetry. Unlike the full string theory,
which is finite, the truncated theory has infinities that
require renormalization. We interpret \cite{emnqm,emnuniv}
the renormalization scale
as the time variable, and renormalization group flow in the
sense of Zamolodchikov \cite{zam}
provides an arrow of time. Non-zero
renormalization coefficients associated with space-time
background fluctuations contribute to the $\nd{\delta H}$ term in
the modified Liouville equation (\ref{modliouv}):
\be
 \delta \nd{H}=\beta ^i  G_{ij}
 \frac{\partial \rho}{\partial p_i}
\label{string}
\ee
where $G_{ij} =  <V_i V_j > $ denotes the
Zamolodchikov metric
in coupling constant space. This extra term
causes a monotonic increase in entropy:
\be
 \dot S= \beta^i G_{ij} \beta^j S
\label{entropy}
\ee
and the asymptotic suppression of off-diagonal elements in the
density matrix \cite{emohn,emnqm}. These
results indicate strongly that the scattering
of light particles is described in string theory by a
non-factorizing $\nd{S}$ matrix, and the present paper
exhibits explicit non-trivial contributions to the $\nd{S}$ matrix.
\pr

We base our discussion on
the Wess-Zumino coset
model \cite{witt}
that characterizes a spherically-symmetric black hole in
string theory. This is an {\it illustration} of the structures
that appear in space-time foam in string theory. We use it
as a convenient tool because it is exactly solvable.
We discuss two types of topological
fluctuations on the world-sheet: monopole-antimonopole
pairs \cite{emndua,ovrut}
corresponding to the quantum creation and annihilation
of a black hole \cite{emndua}, and instantons \cite{yung}
in the Wess-Zumino
coset model.
As we show in section 2,
these instantons describe quantum jumps between
black holes of different masses. Both
monopoles and instantons
increase the central charge
of the string, making it {\it locally}
supercritical with $c  > 26$.
This would be a general feature of fluctuations
in the space-time foam.
We show
that instantons rescale the metric in the same way as one of the
marginal operators associated with massive string modes. Thus
instantons represent the back-reaction expressed by the $W$-induced
coupling of light particles to massive string states.
\pr
As we discuss in section 3, the transition between initial-
and final-state density matrices is given by an absorptive part
of a world-sheet correlation function with an even number of
external tachyon fields, reminiscent
of the Mueller \cite{mueller}
description
of inclusive reactions in hadronic physics. We argue that, as in
the analysis of non-perturbative electroweak cross-sections, there
are identifiable
non-factorizable contributions to the correlation
function, and hence the $\nd{S}$ matrix, from monopole-antimonopole
and
instanton-anti-instanton configurations. Furthermore, the
dominant transitory space-time may correspond to
a valley trajectory \cite{yung1,khoze}.
\pr
It is generally believed \cite{dorinst}
that the motion along a valley
trajectory is equivalent to the scattering of
solitons in one higher dimension. In section 4 we use
this equivalence to identify valley trajectories in the
monopole-antimonopole and instanton-anti-instanton systems.
In the case of the monopole-antimonopole valley, breaking of
conformal symmetry at the quantum level plays a key r\^ole.
In the case of the instanton-anti-instanton valley, the
energy integral must first be made finite by introducing
an infrared cut-off.
\pr
As we discuss in section 5, the
ultraviolet renormalization scale can be identified
with the Liouville field, which can in turn
be identified with target time in a supercritical
string theory \cite{aben}, which is our case. The
arrow of time corresponds to black hole decay.
\pr
We discuss first
in section 6 string contributions to the conventional
$S$ matrix within our renormalization group approach,
associated with topologically-trivial space-time backgrounds.
Then we exhibit the valley contributions to the $\nd{S}$
matrix, and show that they are renormalization scale-, and hence
time-, -dependent, and therefore make contributions
to $\nd{\delta H}$. By analogy with conventional models of
quantum friction \cite{vernon,cald}, these
contributions suppress off-diagonal
density matrix elements. This can be understood intuitively from
the fact that
the
size of the string world-sheet in target space shrinks as
time increases. This means that final-state fields that are
located at arbitrary points on the world-sheet are asymptotically
coincident in target space. Thus off-diagonal terms in the
final-state density matrix in configuration space, which represent
interferences between objects at different points in space,
vanish asymptotically at large times \cite{emohn,emnqm}. This
partial
``collapse of the wave
function" is more rapid for systems containing larger numbers
of particles \cite{emohn2,emnqm}. The final state is
described probabilistically: Schroedinger's cat is either dead
or alive, but not in a superposition of the two states.
\pr
Finally, in section 7 we relate our work to previous
ideas \cite{bohm,penrose} about
wave function collase and the possible r\^ole
of quantum gravity.
\pr
\section{Stringy Black Holes and Instantons}
\pr
In string theory, spherically-symmetric black holes, which are
equivalent for our purposes
to black holes
with a two-dimensional
target space, are described by a gauged SL(2,R)/U(1) Wess-Zumino
coset model \cite{witt}. The black hole exists in both Minkowski and
Euclidean versions, obtained by gauging different subgroups of
SL(2,R). As usual in instanton calculations, it is more
convenient to work with the Euclidean version, and make an
analytic continuation at the end of the calculation.
The Euclidean stringy black hole is obtained by gauging the
axial subgroup of SL(2,R): after eliminating the U(1) gauge field
one arrives at the action
\be
S=\frac{k}{4\pi} \int d^2z [(\partial _\mu r)^2
+ tanhr^2 (\partial_\mu \theta )^2 + \dots    ]
\label{Yung 8}
\ee
where $r$ and $\theta$ are radial and angular coordinates respectively,
and the $\dots$ denote dilaton terms that arise
from the path-integral measure of the gauge-field.
It is convenient to rewrite (\ref{Yung 8}) using the complex
coordinate
\bea
w=sinhr e^{-i\theta} \nn \\
{\bar w}=sinhr e^{i\theta}
\label{yung10}
\eea
in terms of which the action (\ref{Yung 8}) becomes
\be
S=\frac{k}{4\pi} \int d^2z \frac{1}{1+|w|^2}\partial _\mu {\bar w}
\partial ^\mu w + \dots
\label{Yung 11}
\ee
whence we see that the target space-time
line element is
\be
ds^2=\frac{dwd{\overline w}}{1 + w{\overline w}}=dr^2+tanh^2rd\theta^2
\label{metr}
\ee

\nk{\it World-Sheet Vortices/Monopoles}
\pr
As discussed in ref. \cite{emndua}, the Euclidean
black hole (\ref{Yung 8}, \ref{yung10}, \ref{Yung 11}, \ref{metr})
can be written as a world-sheet vortex-antivortex pair,
which is a solution $X_v$ of Green function equations
on a spherical world-sheet of the type
\be
\partial _z\partial _{\bar z} X_v =i\pi \frac{q_v}{2}
[\delta (z-z_1)-\delta(z-z_2)]
\label{vortex}
\ee
where $z_1$ is the location of the vortex and $z_2$ the location of the
anti-vortex (the net vorticity is always zero on a compact world-sheet).
If the vortex is located at the origin and the antivortex at
infinity (corresponding to the South and North Poles in a
stereographic projection), the corresponding $\sigma$-model
coordinate solution of (\ref{vortex}) is
\be
X_v=q_vImln z
\label{vortsol}
\ee
It is clear that for $X_v$ to be angle-valued
it must have period $2\pi$, and hence the vortex
charge $q_v$ must be an integer.
\pr
To see the interpretation of (\ref{vortsol}) as a
Euclidean black hole \cite{emndua},
we rewrite the vortex configuration
$X_v \equiv \theta $ as
\be
e^{2i\theta} = \frac{z}{\bar z}
\label{vorexp}
\ee
and complexify the phase by introducing a real part $r$, defined
by the following embedding of the world-sheet in a two-dimensional
target space ($r$,$\theta$):
\be
z=(e^r -e^{-r})e^{i\theta}
\label{embed}
\ee
The induced target-space metric (\ref{metr}) is inferred from
the world-sheet arc-length, which is $dl=\frac{dz}{1+z{\overline z}}$,
after the stereographic projection, and the infinitesimal
Euclidean displacement $d{\bar z}$ induced by a corresponding
shift in the
$r$-coordinate.
\pr
The representation of the Euclidean black hole as a coset
$SL(2,R)/U(1)$ Wess-Zumino model means that
the world-sheet vortices (\ref{vortex}) induce gauge
defects of the underlying world-sheet gauge theory. Eliminating
the non-propagating gauge field $A_z$ of the gauged Wess-Zumino model
via its equations of motion yields
\be
A_z=-\frac{u\partial _z (a-b)-(a-b)\partial _z u}{(a+b)^2}
\label{defect}
\ee
where $u=sinhr sin\theta$, $a=coshr + sinhr cos\theta $ and
$b= coshr -sinhr sin\theta$. In the neighbourhood of the
singularity $r\rightarrow \epsilon$, $u\rightarrow \epsilon sin\theta$,
and $a-b \rightarrow 2\epsilon cos\theta$, so that the gauge
potential (\ref{defect}) becomes that of a singular
gauge transformation
\be
A_z \rightarrow \epsilon ^2 \partial _z\theta
\label{sing}
\ee
which has
the interpretation of a $U(1)$ gauge
monopole \cite{emndua}.
\pr
In addition to vortex configurations, one can
have
monopole-antimonopole (or better,
``spike-anti-spike'')
pairs
on the world-sheet.
These are
solutions $X_m$ of Green function equations of
the following type \cite{ovrut}:
\be
   \partial _z  {\overline \partial _z} X_m =-\frac{q_m \pi} {2}
[\delta (z-z_1) -\delta (z-z_2)]
\label{monpair}
\ee

\nk which corresponds to a ``spike''
at $z_1$ and an ``antispike'' at $z_2$.
Once again, the compactness of the closed string world-sheet imposes
zero net ``spikiness'', but in this case $X_m$ is non-compact and
hence aperiodic, so the spike charge $q_m$ is not quantized $a$
$priori$.
After
stereographic projection of the South Pole on the
sphere onto the origin and of the North Pole onto the point at infinity
in the complex plane,
\be
  X_m = q_m Relnz=q_m ln|z|
\label{monop}
\ee
The world-sheet
monopoles are related to Minkowski black holes \cite{emndua},
and the corresponding charges are proportional to the black
hole mass.
\pr
Both vortex and spike
configurations can
be described in terms of sine-Gordon
deformations of the Lagrangian for the field $X$.
Viewing these defects as thermal excitations on the
world sheet, corresponding to a `pseudo-temperature' $\beta^{-1}$,
and
assuming a statistical population,
the corresponding effective action reads \cite{ovrut}
\be
Z=\int D{\tilde X} exp(-\beta S_{eff}({\tilde X}) )
\label{act}
\ee

\nk where ${\tilde X} \equiv   \beta^{\frac{1}{2}}X$, and
\bea
\nonumber
\beta S_{eff} &=& \int d^2 z [ 2\partial {\tilde X}
{\overline \partial } {\tilde X} +  \frac{1}{4\pi }
[ \delta _v\omega ^{\frac{\alpha}{2}-2}
(2 \sqrt{|g(z)|})^{1-\frac{\alpha}{4}}: cos (\sqrt{2\pi \alpha }
[{\tilde X}(z) + {\tilde X}({\bar z})]):   \\
&+&  (\delta _v, \alpha,
{\tilde X}(z) + {\tilde X}({\bar z}) )
\rightarrow (
\delta _m, \alpha ', {\tilde X}(z) - {\tilde X}({\bar z}))]]
\label{eff}
\eea

\nk Above we have made a stereographic projection of the sphere of radius
$R$ onto the complex plane, inducing an effective metric $g(z)$;
$\omega$ is an angular ultraviolet
cut-off on the world-sheet, which accompanies
the normal ordering of the cosine (deformation) terms.
For future use, we note that the projection of the
cut-off on the
sphere of radius $R$ is $2\omega R$.
Its stereographic
projection on the plane through a point $z$, as in figure 1,
yields
$2 \omega R (1 + |z|^2/4R^2) $. In the Liouville
theory framework, that we follow in this paper,
we can keep the radius of the sphere
{\it fixed} and let $\omega$ vary, which thus incorporates
the effects of the Liouville mode. Equivalently,
one may assume a spherical contraction of the world sheet
(as the theory approaches close to the ultraviolet
fixed point), keeping $\omega$ fixed.
The quantities
$\delta_{v,m}$ in eq. (\ref{eff})
are the fugacities for vortices and spikes
respectively, and
\be
 \alpha \equiv 2\pi \beta q_v^2  \qquad \alpha ' \equiv
2\pi \beta q_m^2
\label{anom}
\ee

\nk are related to the conformal
dimensions $\Delta_{v,m}$ of the vortex and
spike creation operators respectively, namely
\bea
\nonumber
\alpha =4 \Delta _v \qquad   \alpha ' =4 \Delta _m \\
      \Delta _m =\frac{\pi\beta}{2}q_m^2=
      \frac{(eq_v)^2 }{16 \Delta _v}
\label{conf}
\eea
where we took into account the vortex-induced
quantization condition for monopoles \cite{ovrut}
$2\pi \beta q_m =e  :  e=1,2,\dots $
The corresponding
deformations
are irrelevant - and the
pertinent
topological defects are thus bound in pairs -  if the
conformal dimensions are larger than one.
\pr
For future use we mention that
these considerations have also been extended to
Liouville theory \cite{ovrut}.
If $\phi$ is a Liouville mode, the corresponding
action on a curved world-sheet with metric $g_{ab}$
reads
\be
S_L=(\frac{25-c}{96\pi})\int d^2 x \sqrt{g}(g^{ab}\partial _a \phi
\partial _b \phi + \phi R^{(2)})
\label{liouv}
\ee
where $c$ is the central charge of the matter theory.
For $c \ge 25$ the requirement of the boundedness
of the action (\ref{liouv}) implies a Wick rotated, angle-valued
Liouville field $\phi$, and the theory
admits  both
monopoles and vortices. The r\^ole of the
pseudo-temperature
is now played by the quantity \cite{ovrut}
\be
\beta _L=\frac{3}{\pi (c-25)}
\label{pstl}
\ee
An interesting consequence of (\ref{pstl}), which will be of use to us
later, is that for $e=1$
the monopole operator is irrelevant
in the region where $c \ge 49$.
In this region the vortex operator is relevant.
In fact, in general for $c \ge 25$ there is no region where both
operators are irrelevant, and thus the system
appears strongly coupled. We also note for future reference
that, in accordance
with the Zamolodchikov c-theorem \cite{zam},
the part of the central charge
pertaining to the conformal field theory subsystem that possesses the
topological defects increases when the latter appear
bound in dipole-like pairs as irrelevant deformations.
\pr

\nk{\it Instantons}
\pr
These appear \cite{yung}
in the $SL(2,R)/U(1)$ Wess-Zumino
coset model as follows. A topological charge is defined
in this model
by analogy with
compact $\sigma$ models:
\be
 Q=\frac{1}{\pi}\int d^2z \frac{1}{1+|w|^2}[{\overline \partial}
{\overline w}\partial      w  - h.c. ]
\label{Yung 15}
\ee
It is easy to verify that there are topological classes labelled by an
integer $n$, and that
the action is minimized in each
topological class by holomorphic functions
\be
 w(z)= b    \frac{(z-c_1)...(z-c_n)}{(z-z_1)...(z-z_n)}
\label{Yung 19}
\ee
where the parameters ${ b, c_{1,2,...,n}, z_{1,2,...,n}}$ are
complex, and $n$ is the winding number of the map (\ref{Yung 19}).
The topological charge (\ref{Yung 15})
is proportional to the
winding number $n$, but with a logarithmically-divergent
coefficient:
\be
 Q=-2nln(a) + ~const
\label{Yung 21}
\ee
where $a$ is the ultraviolet cutoff (in configuration space)
to be discussed in more
detail later\footnote{To make contact with the
case of spherical world sheets, discussed above, we
remind the reader that this cut-off is $2\omega R$,
in the notation above.}. One should stress at this point that the
black hole metric $g(|w|)=(1+|w|^2)^{-1}$
is the limiting case allowing for the existence of
instanton solutions, as it leads to logarithmic,
and not to power,
dependence of the topological charge $Q$
on the cutoff.
The coefficient of the leading logarithm in $Q$ is
regularization-independent and hence the quantity $Q$,
although
divergent, still allows for a definition of a
winding number $n$ and, therefore, of topological classes
for the configuration space of the model.
\pr
We concentrate here on the simplest
$n$ = 1 instanton which can be written as
\be
 w(z)=\frac{\rho}{z-z_0}
\label{Yung 23}
\ee
where $\rho$ is the instanton size parameter. The contribution of
this holomorphic instanton and its antiholomorphic anti-instanton
partner to the free energy of the model is given in the dilute gas
approximation by
\be
 F^{dilute-gas} \simeq -\frac{d}{2\pi}\int d^2z_0 \frac{d^2\rho}
{|\rho|^4} e^{-S_0}
\label{Yung 24}
\ee
where $d$ is an appropriate constant and
$S_0$ is the action of the instanton
(\ref{Yung 23}), which is
easily evaluated to be
\be
 S_0 = \frac{k}{2}ln(a^{-2}|\rho|^2 + 1)
\label{instact}
\ee
Inserting this result into the formula (\ref{Yung 24})
for the free
energy, we
find a decrease in the free energy:
\be
 F^{dilute-gas} \simeq -d\int d^2z_0
\frac{d|\rho|}{|\rho|^3}(a^{-2}|\rho |^2 +1)^{-\frac{k}{2}}
\equiv -d'a^{-2}V^{(2)}
\label{Yung 27}
\ee
where
$V^{(2)}$ is the (two-dimensional) world-sheet
volume. The constant $d'$ is the instanton density.
We shall discuss it in more detail later on.
The effect of the instanton (\ref{Yung 23})
can be reproduced by
the vertex operator
\be
V_I \propto \int d^2z \frac{d^2\rho}{|\rho|^4}
e^{-S_0}e^{k[\rho \partial {\overline w} + h.c.]}
\label{Yung 28}
\ee
which can be combined with the corresponding anti-instanton
vertex to give
\be
V_{I{\overline I}}\propto -\frac{d}{2\pi}
\int d^2z \frac{d^2\rho}{|\rho|^4}
e^{-S_0} (e^{(\frac{k[\rho\partial {\overline w} + h.c. + \dots ]}
{f(|w|)}}+ e^{(\frac{k[\rho \partial w + h.c. + \dots]}{f(|w|)}})
\label{Yung 33}
\ee
where $f(|w|)$ is an unknown function that is regular at $w  = 0$.
It can be expandend as  $f(|w|) = 1 + a_1 |w|^2 + a_2|w|^4 + \dots$.
In the dilute-gas approximation it is not possible to
compute this function. It is conceivable that
$f(|w|)$ is the inverse metric $(1+|w|^2)^{-1}$. In this
case the {\it exact}
effects of instantons would be correctly
captured by the large-$k$ expansion, as we discuss immediately
below.
\pr
At leading order in large $k$ we need only retain the first two terms
in a power series expansion of $f (| w |)$ about the origin, and
the effect of the instanton vertex
\be
 V_{I{\overline I}} =-d' [ a^{-2}V^{(2)} +\frac{k^2}{2}
\int d^2z \frac{\partial_\mu{\overline w}\partial^\mu w}{f^2(|w|)} ]
\label{Yung 37}
\ee
is simply to add a vacuum energy term, which we discuss later,
and to renormalize the kinetic energy term of the coset model
(\ref{Yung 11}):
\be
 k \rightarrow k - 2\pi k^2 d'
\label{Yung 38}
\ee
In the above expression $d'$ denotes the instanton density
which is given via
 (\ref{Yung 27}) as
\be
 d' \equiv d\int \frac{d|\rho|}{|\rho|^3}
\frac{a^{2}}{[(\rho/a)^2 + 1]^{\frac{k}{2}}}
\label{density}
\ee
The integral (\ref{density}) has a power divergence
for small-size instantons ($\rho \rightarrow 0$), and
one is
obliged to cut
such contributions off
at a scale $a$.
Such
a procedure results in an effective renormalization of
the Wess-Zumino level parameter $k$. Indeed,
consider the right-hand-side
of the expression (\ref{density})
integrated over small-size
instantons only, i.e.
$d'=d \int_{0}^{1} d{\hat \rho}\frac{1}{{\hat \rho}^3
(1+{\hat \rho}^2)^{\frac{k}{2}}}$, and concentrate in the region
${\hat \rho} \equiv \frac{\rho}{a}
 << 1$. In this region we can parametrize
${\hat \rho}\equiv
|\frac{a}{\Lambda}|^{\gamma}$, where $\Lambda$ is
the world-sheet
infrared cut-off in configuration space,
and $\gamma$ a small positive number. Converting the integration
over ${\hat \rho}$ into an integral over $\gamma $
around a small value $\gamma_0$, of radius
$\simeq \frac{1}{|ln|a/\Lambda||}$ (which is the region corresponding
to the leading ultraviolet divergences)
we get, in the limit $a        \rightarrow 0$
\be
 d'\simeq |\frac{a}{\Lambda}|^{-2\gamma_0}
\label{scaledepd}
\ee
It can be shown that if
one considers matter
deformations in the theory, as we shall do later on, the relevant
order of magnitude of $\gamma _0$ is, to leading approximation
in
the deformation couplings, of the order of the anomalous dimension
of the matter deformation.
The induced
divergences can be absorbed,
as a result of (\ref{Yung 38})
in a `renormalization'
of $k$, which, thus, becomes scale-dependent.
It should be stressed that the above computation
is only {\it indicative}, in order to demonstrate
the scale dependence of $d'$ due to regularization.
The anomalous dimension
coefficient $\gamma _0$ is assumed small enough
so as to guarantee the validity of a perturbative
$\epsilon$-expansion when discussing
contributions to the light-matter
$\nd{S}$ matrix, as we shall see in section 6.
There we will extrapolate the results
to finite $\gamma _0$, analogously to
the $\epsilon$-expansion in local field theory,
whose qualitative
features are assumed valid in the region of finite $\epsilon$.

\pr
Performing the above regularization procedure, and
rescaling
$\rho \rightarrow     \rho/a $,
we arrive easily, after a change of integration variable
$|\rho/a| \rightarrow \tau=\frac{1}{|\rho/a|^2}$, at
\be
d_{reg}'=\frac{d}{2}\int _{0}^{1}d\tau (\tau)^{\frac{k}{2}}
(1+\tau)^{-\frac{k}{2}} =\frac{d}{2(\frac{k}{2} + 1)}
2^{-\frac{k}{2}}
F(1,\frac{k}{2}, \frac{k}{2}+2 ; \frac{1}{2})
\label{cprime}
\ee
The series expansion of the
hypergeometric
function yields terms of the form \cite{abram}
\bea
F(1,\frac{k}{2}, \frac{k}{2}+2 ; \frac{1}{2}) =
\frac{1}{2}\Gamma (\frac{k}{2}+1)+ \{
\frac{1}{2}(\frac{k}{2} +1)\frac{k}{2} \times
\nn \\
\times
\sum _{n=0}^{\infty}\frac{(2)_n(\frac{k}{2} +1)_n}{n!(n+1)!}(2)^{-n}
[ln(1/2)-\psi(n+1)+\psi(\frac{k}{2}+n+1)] \}
\label{series}
\eea
Hence in the large-$k$ limit
$d'(k) << 1$ and the
convergence of the  expression (\ref{Yung 38}) is guaranteed.
\pr
Since the level parameter $k$ is inversely related to the central
charge $c$
\be
c=\frac{2(k+1)}{k-2}=2 + \frac{6}{k-2}
\label{central}
\ee
the effect of the decrease (\ref{Yung 38}) in $k$ is to increase
the central charge. If we assume that the stringy black hole
had a space-time interpretation before adding instantons, so
that its central charge was 26, either because $k$ = 9/4 or
because additional matter fields were present, the string model
will be noncritical after including instanton effects. Indeed,
it will be supercritical with $c  > 26$, reflecting the fact
that the instanton vertex is an irrelevant operator.
As already commented in section 1, the fact that
both monopoles and instantons make the string
{\it locally} supercritical is presumably
a general feature of incorporating foamy backgrounds
in string theory.
\pr
As is apparent from the representation (\ref{Yung 8}) of the stringy
black hole action and the form (\ref{Yung 33}) of the instanton vertex,
the shift (\ref{Yung 38}) in the level parameter
$k$ renormalizes the target space metric. This rescaling is
correlated with the renormalization of the black hole mass:
\be
 M_{bh}=\sqrt{\frac{2}{k-2}}e^{const}
\label{mass}
\ee
The effect of the instanton, at least as seen
in the large $k$ limit \footnote{We note once again
that this large-$k$ effect would
be exact if the function
$f(|w|)$ were the inverse target-space metric.}, is
therefore to increase the black
hole mass in a quantum jump which may represent back-reaction or
infall. Its effect is $opposite$ to that of higher genera
(loops) \cite{emnloop},
which $increase$ $k$, and hence $decrease$ $c$, corresponding to
black hole decay. Changes in the stringy black hole background
contribution to the total central charge can always be compensated
by a change in the contribution of other, matter, fields. This
reflects the fact that operators which appear irrelevant within a
specific conformal model may become marginal in the context of
models with a larger value of the central charge.
\pr
In what follows, we would like to
make a connection of these results
with the $SL(2,R)$ current algebra
deformation approach of ref. \cite{chaudh}.
We
shall be interested in exactly
marginal deformations of conformal string
backgrounds.
Such deformations are (1,1) operators which however
retain their conformal dimension in the deformed theory, thereby
generating families of conformally invariant backgrounds.
Such operators
in the Wess-Zumino coset model describing the
stringy black hole have been studied in \cite{chaudh}. They
consist in general of vertices for both massless (tachyon) and
massive states. Since the latter are solitons in the stringy
background, these exactly
marginal operators represent the back-reaction
of matter fields on the metric. One particular example of such an
exactly
marginal operator is \cite{chaudh}
\be
  L_0^2{\overline L}_0^2 \propto
  \psi^{++} + \psi^{--}  + \psi^{-+} + \psi^{-+} +
\dots
\label{marginal}
\ee
where
$\psi^{\pm\pm} \equiv :
({\overline J}^{\pm})^N(J^{\pm})^N(g_{\pm\pm})^{j+m-N}: $,
$\psi^{\pm\mp} \equiv :
({\overline J}^{\pm})^N(J^{\mp})^N(g_{\pm\pm})^{j+m-N}:$,
are composite fields constructed out
of the $SL(2,R)$ currents $J$
and the
Wess-Zumino field $g$, and represent {\it massive}
discrete string states
at level $N$.
The $\dots$ in eq. (\ref{marginal})
imply summation over {\it all} such states.
We see immediately \cite{chaudh}
that this operator also
rescales the target space metric.
In the context of the critical $k=9/4$
string theory,
such global rescalings of the metric
define families of conformally invariant black-hole
backgrounds, which may be thought of as corresponding to
constant shifts in the dilaton field \cite{witt}.
{}From a target-space point of view, however, such global
rescalings of the metric field amount to a change in the
black hole mass (\ref{mass})
which might also be attributed
to a change in $k$ analogously to (\ref{Yung 37}).
\pr
To understand this interpretation better, we
give another example of an exactly marginal
deformation of the Wess-Zunino stringy black hole.
It assumes the form:
\be
L_0^1{\overline L}_0^1 \propto
\Phi ^{c-c}_{\frac{1}{2},0,0} + i(\psi^{++}-\psi^{--})) +\dots
\label{margintax}
\ee

\nk where the massive string modes $i(\psi ^{++}-\psi^{--})+ \dots$
are given
in terms of $SL(2,R)$ currents as in (\ref{marginal})
\cite{chaudh}. The operator
$\Phi ^{c-c}_{\frac{1}{2},0,0}(r)$
generates
the light string matter.  The level-one massive modes
indicated explicitly
in (\ref{margintax}) represent back-reaction
on the space-time metric, as can be checked explicitly
in the case $k \rightarrow \infty$ \cite{chaudh}.
The  deformation
$\Phi ^{c-c}_{\frac{1}{2},0,0}(r)$
alone is relevant.
Thus, the {\it combined} deformation consisting of
(\ref{marginal}) and the {\it massive} mode (irrelevant)
parts of (\ref{margintax}) is an {\it irrelevant}
deformation of the Wess-Zumino theory, which
represents back-reaction effects of matter on the space-time
black-hole geometry. Such effects are analogous to those
induced by instantons (\ref{Yung 37}).
Although at this stage this analogy cannot be made
rigorous, however it follows that
the instanton can represent the effects
of massive string modes\footnote{These
are related to each other and to the massless excitations by
a particular quantum version of the $W_{1+{\infty}}$ algebra,
with the level parameter $k$ interpreted as a quantum deformation
parameter.}.
\pr
There is a formal way of demonstrating this,
to lowest order in perturbation theory,
by
looking at the behaviour of the model with instantons
under renormalization group flow.
We first observe that, due to (\ref{instact}),
the dominant contributions
to the path-integral
in the dilute gas approximation come from instantons with sizes
of the order of the cut-off. This is expected because
the instanton perturbations (\ref{Yung 33})
are irrelevant operators
in a renormalization group sense. To see this, we
first note that on dimensional grounds
one expects the instanton contributions to be of order
\be
(\rho/a)^{2\lambda}
\label{incontr}
\ee

\nk where $\lambda$ is the first
coefficient of the $\beta$-function. In our case the instanton
$\beta$ function can be estimated by looking at the contribution
of the deformation (\ref{Yung 33}) to the trace of the
stress-tensor of the model. The result is \cite{yung}
\be
 \beta=-\frac{k}{2}
\label{betainst}
\ee

\nk which is negative for $k > 0$, hence the
irrelevance of the respective deformation.
The result (\ref{betainst}),(\ref{incontr})
is compatible with (\ref{instact}). This result implies that
the dominant contriutions in the path-integral
are instantons of the size of the cut-off
\be
 \rho \simeq a
\label{dominant}
\ee
To discuss the r\^ ole
of contributions close to the instanton
center, it is convenient to
keep $\rho$ fixed and finite
in a correlation function
involving $N$ tachyon
operators $T_i \equiv T(z_i)$, while we send $|z_i-z_0| \rightarrow 0$
(i=1,2,...N).
One may
represent such a process
as a sum of zero and non-zero instanton sectors appropriately
normalized. From the general analysis in the zero-instanton sector
\cite{chaudh} one may represent the tachyon deformations
in the $\sigma$-model as
\be
\Phi ^{c-c}_{\frac{1}{2},0,0}(r)
=\frac{1}{coshr}
F(\frac{1}{2},\frac{1}{2} ; 1, tanh^2r )
\label{tachyon}
\ee

\nk where $r$ is a target spatial coordinate
defined in (\ref{Yung 8}).
\pr
In the region of weak matter deformations, $r $ is large and one
may use the following expansion of the hypergeometric
function \cite{abram}
\bea
&~&F(\frac{1}{2},\frac{1}{2},1;tanh^2r) \simeq
\frac{1}{\Gamma ^2(\frac{1}{2})}\sum_{n=0}^{\infty}
\frac{(\frac{1}{2})_n(\frac{1}{2})_n}{(n !)^2}[2\psi(n+1)-
2\psi(n+\frac{1}{2})+ \nn \\
&+&ln(1 + |w|^2)]
(\sqrt{1 + |w|^2}~)^{-n}
\label{expansion}
\eea

\nk The leading behaviour for $r$ large
comes from the terms with
$n=0$. In this limit
the tachyon function yields the
usual result of the tachyon background \cite{tachback}
\be
T(r) = sech(r)ln(cosh(r))
\label{tacbac}
\ee
which after an appropriate change of variable becomes identical to the
result
of the perturbative $\beta$-function
approach to
$c=1$ string theory \cite{witt,alwis}.
In this black hole string theory, however,
the
tachyon operator (\ref{tachyon}) alone
is no longer a marginal
deformation in the presence of a black hole, but a relevant
deformation,
since the
massive string mode effects in the truly marginal operator
(\ref{margintax}) are not vanishing.
This
non-vanishing of the $\beta$-functions of the light string modes
is  the key ingredient to define target time as an evolution
parameter in a renormalization-group sense \cite{emnqm}.
If we denote generically the `bare' couplings of the tachyon deformations
by $T^i$ then, to leading order, the renormalized couplings are
of the form
\be
T_R^i \propto|\frac{a}{\Lambda}|^{\lambda}T^i
\label{renorm}
\ee

\nk where $\Lambda$ is the infrared
cut-off introduced previously
for dimensional reasons\footnote{Here we follow
 Wilson's renormalization approach, where  the cut-off is never
removed and the variable scale is the ratio of infrared to ultraviolet
cut-offs. It is this approach that was used in \cite{santos}
to prove the gradient flow of the renormalization group $\beta$-functions
used in the equivalence of the $\sigma$-model
conformal conditions with
the $S$-matrix approach to string theory \cite{mavmir}.
In the Wilson approach, elimination of the irrelevant operators
leads eventually to the perturbative Gell-Mann-Low $\beta$-functions,
that do not have an explicit cut-off dependence.}.
\pr
To demonstrate
things clearly we consider first the one-point tachyon correlator
in the one-instanton background. This means that we evaluate
$<V_T V_{inst}>T^i$, where $V_T$ is the vertex operator for the tachyon
given in (\ref{tachyon}), and the instanton vertex is given by
(\ref{Yung 33}).
In the coset model (\ref{Yung 11}) the following relation
is valid to leading order in $k$
\cite{yung}:
\be
\partial _z < w(z_1)w(z_2)> \propto \frac{1}{k} \frac{1}{z_1 -z_2 }
\label{relation}
\ee

\nk which implies that when evaluating correlation functions
of $w$ in the one-instanton sector for large $k$
one effectively replaces $w$
by its classical expression (\ref{Yung 23}),
and its complex conjugate for anti-instantons.
Taking into account the expansion (\ref{expansion})
of the hypergeometric
function in (\ref{tachyon}), we find for
the one-point function
in the short-distance region of the $z$-integration
terms proportional to

\be
T_R (\frac{|a|}{\Lambda})^{-\lambda}
 \int _{z \rightarrow z_0}
 d^2 z  d^2z_0 d^2\rho \frac{1}{|\rho |^4}
(\frac{|\rho|^2}{|a|^2}+1)^{-\frac{k}{2}}
\frac{|z-z_0|}{|\rho |}2ln|\frac{|\rho|}{|z-z_0|}
\label{instrenorm}
\ee

\nk Recalling
that the dominant contributions
to this integral come from instantons with sizes
of order $a$, and cutting off the instanton
size integration
at $a$, we observe that the tachyon deformation
(\ref{tacbac})
is marginal in the one-instanton sector,
in the sense that the integrand ({\ref{instrenorm})
has a smooth short distance behaviour.
%
%
This implies a scale-independent $T$ in this sector, because the
anomalous dimension induced by the instantons cancels that
of the ordinary renormalization of $T$ in the zero-instanton sector.
This is exactly what should happen in the dilute gas approximation,
given the argued r\^ ole of the instantons as representing
massive string mode effects. The combined  effect of the (exact)
instanton and light matter deformations is expected to be
marginal, and the above analysis checks
this conjecture
in the single instanton sector for large $k$.
\pr
\section{Topologically Non-Trivial
Contributions to Transitions between Density
Matrices}
\pr
In string theory,
all physical quantities can be related to the calculation of
correlation functions on the world-sheet. In our case, these
are to be
evaluated in a black hole background in target space,
including the effects of world-sheet monopoles or
instantons. In the dilute gas
approximation, correlation functions for light matter (tachyon)
fields may be expanded in powers of the
density of topological structures:
\be
\frac{<A|B>_{inst}}{<0|0>_{inst}}=
\frac{<A|B>_0}{<0|0>_0}+\frac{<A|B>_1}{<0|0>_0}-
\frac{<A|B>_0}{<0|0>_0}\frac{<0|0>_1}{<0|0>_0} +\dots
\label{corrfn}
\ee
where $<\dots>_{1(0)}$ denotes correlators evaluated
in the
one(zero)-monopole or-instanton
sector. From the discussion in the previous section, we see that
the zero-monopole or -instanton
contributions
correspond to neglecting the
back-reaction of matter fields on the metric. These constitute
the approximation made up to now in the string treatment of
black hole physics. There is a well-defined $S$ matrix for the
scattering of light particles off a black hole in this
approximation, whose only non-zero elements are those with
just one incoming or outgoing tachyon \cite{klepol,emnsel}:
\be
G_{-++...+}, G_{+--...-} \ne  0
\label{matrcorr}
\ee
whilst all other Green functions vanish.
\pr

We wish to calculate the first non-trivial
corrections to this result, corresponding to the first
non-trivial terms
in the
expansion (\ref{corrfn}). Rather than
calculate this correction for all Green functions, we adapt a
convenient approach used recently \cite{khoze}
to estimate non-perturbative
effects in electroweak theory, which exploits the optical theorem.
\pr
We are interested in the transition between a generic initial-state
density matrix ${{\rho^A}_B}^{in}$ and final-state density matrix
${{\rho^C}_D}^{out}$. Here the index $A$ labels a generic bra vector
$< A | _{ in }$, and similarly for the other indices: $| B > _{ in }$,
$< C | _{ out }$ and $| D > _{ out }$. These states may each contain
many matter fields at different target-space locations. If not
stated explicitly, we will consider the indices $A$, etc., as
referring to the target-space locations. In string theory,
these are coordinates given by the values of field variables on the
world-sheet: $X$($z_A$) = A, etc..
We note that the field for a single light particle
(`tachyon') may involve an infinite number of world-sheet
operators at the same world-sheet point. For example,
far from the centre of the string black hole, in the weak field
region, one may Fourier expand a tachyon field:
\be
T(X)=\int d^Dk e^{ik_\mu X^\mu}T(k)
\label{fourier}
\ee
where $T(k)$ is a `tachyon' polarization tensor in momentum
space.
In the stringy black hole, there is
a one-to-one mapping between points in target space and on the
world-sheet: $A \leftarrow\rightarrow z_A$,
etc..
In general, transition matrix
elements are calculated from matter (tachyon) correlation
functions on the world-sheet. In the simplest case of single-particle
initial and final states, the relevant object is a four-point
correlation function. The quantity required is a specific
absorptive part of this correlation function, analogous to the
discontinuity that describes \cite{mueller}
an inclusive cross-section in
conventional hadronic physics:
\bea
&~&\sum _{X_{out}}~_{in}<A|D,X>_{out}~_{out}<X,C|B>_{in} =  \nn \\
&=&\sum _{X_{out}}~_{in}<0|T(\phi(z_A)\phi(z_D))|X>_{out}
{}~_{out}<X|{\overline T}(\phi(z_C)\phi(z_B))|0>_{in} = \nn \\
&=&~_{in}<0|T(\phi(z_A)\phi(z_D))
{\overline T}(\phi(z_C)\phi(z_B)|0>_{in}
\label{scattr}
\eea
where the symbols $T$ and ${\overline T}$ denote time-ordered
(anti-ordered) products, and equation (\ref{scattr})
has the diagrammatic representation
shown in figure 2.
\pr

Note that we have used the optical theorem
on the world-sheet to replace the sum over unseen states $X$
by unity. This application of the optical theorem is
justified by the fact that conventional quantum mechanics and
quantum field theory remain valid on the world-sheet: it is
only their elevation to space-time that we challenge.
We recall that, in such a hadronic inclusive reaction at high
energies, the final-state particle distribution is described
probabilistically, with all interferences vanishing after
summation over the unseen parts $X$ of the final states.
We will argue later
that something similar happens when one
calculates the analogous absorptive part of the correlation
function on the world-sheet.
\pr

Since the quantity of interest (\ref{scattr})
is an expectation value in
a definite vacuum state, all topologically non-trivial
contributions must have equal numbers of
monopoles and anti-monopoles or
instantons and
anti-instantons.
If the string
theory were critical, the monopole anti-monopole or
instanton-anti-instanton configurations
would be trivial, since they could be moved around arbitrarily
by conformal transformations. However, as discussed in the
previous section, monopoles or
instantons make the stringy black hole
non-critical, so that the action depends on the
relative locations.
The question then arises: what is the dominant
contribution to the functional integral in the
topologically non-trivial
sector? It is
presumably given by a saddle point in the action associated with
a valley trajectory \cite{yung1,khoze},
as in figure 3.
Thus,
to evaluate the absorptive part of a correlation function
in the conventional path-integral
formalism, we first continue analytically
to Euclidean space, and
then make a saddle-point
approximation using a valley configuration. The path integral
is thereby converted into an
integral over the collective coordinates that describe
the valley. The leading semi-classical behaviour of an
$\nd{S}$ matrix element in the single defect-antidefect
valley approximation takes the form
\be
\nd{S} \propto Abs\int D\phi^c exp(-S_v^{(1)}(\phi^c))F_{kin}
\label{valleyap}
\ee
where the integral is over the collective coordinates $\phi^c$
of the valley, whose action is described by $S_v^{(1)}(\phi ^c)$.
The function $F_{kin}$ depends on kinematic factors,
in particular the total centre-of-mass energy
$E$ in the case of a forward four-point
Green function, and on collective coordinates, in
particular the separation parameter $\Delta R$ of the
valley. It is a generic feature of valley
configurations that in the four-point case,
for large $E \Delta R >> 1$,
\be
F_{kin} \simeq exp(E \Delta R)
\label{kinval}
\ee
allowing for a saddle-point approximation to the
integral (\ref{valleyap}).
To evaluate the relevant
absorptive part of (\ref{valleyap})
it is necessary to continue back to
Minkowski space \cite{khoze}.
We assume that, as usual, the
quantitative features of this semi-classical approximation
are not affected by quantum fluctuations.

\section{Valley Trajectories in the String Black Hole }
\pr

We demonstrate in this section the existence in the $SL(2,R)/U(1)$
coset Wess-Zumino model of suitable monopole-antimonopole
and instanton-anti-instanton
valley trajectories, following
the general method outlined in ref. \cite{BW} and applied to the
two-dimensional $O(3)$ $\sigma$-model. Since there are many
similarities, but also some important formal differences, we
first review briefly the case of the $O(3)$ $\sigma$-model, and
then move on to the $SL(2,R)/U(1)$ case.
The reader who is not interested in the details of the
valley trajectories, but is content to accept their
existence
and eager to study their physical consequences,
is advised to go to section 5.
\pr

\nk{\it O(3) $\sigma$-model  valley}
\pr
The "target space-time" metric in the $O(3)$ model is given in
complex notation by $g(w) = 1/(1+|w|^2)^2$, and the action is
\be
S_\sigma = \int d^2z g(|w|)[\partial w {\overline \partial }
{\overline w} + h.c.]
\label{tessena}
\ee
This model has well-known instanton solutions:
\be
w(z)= b    \frac{(z-c_1)\dots (z-c_n)}{(z-z_1)\dots (z-z_N)}
\label{O(3)inst}
\ee
In order for the separation between an instanton and an
anti-instanton to be defined, one must break conformal
invariance, specifically
by adding to the action a term of the form
\be
S_1 =\frac{2m^2}{g^2}\int d^2z \frac{|w|^2}{1+|w|^2}
\label{tessdio}
\ee
When evaluated in the one (anti-)instanton sector, this term yields
\be
S_{I({\overline I})}=\frac{2\pi}{g^2}m^2(\rho _{I({\overline I})}^2)
\label{tesstria}
\ee
and so serves as an infrared cutoff for large-size instantons in
the partition function of the theory. For our purposes, the
importance of this term is that it allows for static finite-energy
solutions that resemble the sphalerons of four-dimensional gauge
theories. Such solutions describe
the passage between topologically-inequivalent
sectors without tunnelling at centre-of-mass energies
comparable to the sphaleron mass, $E_s=\frac{2\pi m^2}{g^2}$,
i.e., the height of the barrier
between different vacua. The sphaleron is the highest-energy
point on the valley trajectory which we now discuss.
\pr

The construction \cite{BW,dorinst}
of the valley trajectory is simplified by
observing that if one restricts one's attention to
radially-symmetric configurations of the field $w =f(r)e^{-in\theta}$
($n$ in $Z^+$), one can map the action (\ref{tessena})
onto that of the
sine-Gordon model by setting $r=e^{y/n}$, $f(r)=tan(\psi/4)$:
\be
S_0=\frac{n\pi}{2g^2}\int dy [\frac{1}{2}   ( \psi ')^2
+ (1-cos\psi)]
\qquad : \qquad (\dots) ' \equiv \frac{\partial}{\partial y}(\dots)
\label{tesstess}
\ee
This transformation maps the instanton solution of the $O(3)$
$\sigma$-model into the kink solution of the sine-Gordon model:
\be
\psi_K (y;y_0)=4tan^{-1}exp(y-y_0)
\label{tersstessp}
\ee
The sine-Gordon model has a kink-antikink valley with
homotopic valley parameter $\mu$ that obeys the equation
\be
\frac{\partial S_0}{\partial \psi}|_{\psi =\psi _v} =
W_\psi(y,\mu)\frac{\partial \psi _v}{\partial \mu}
\label{weight}
\ee
where $W_{\psi}(y,\mu)$ is a weight function,
with the following properties \cite{khoze}: it is
positive definite and decays rapidly at large $y$.
Thus one recovers at large $y$ and/or $\mu$
the asymptotic solution
\be
\psi _v(y;\mu) \rightarrow \psi _K (y, -y_0) + \psi _{\overline K}
(y;y_0)
\label{tessexi}
\ee
Note that the weight function $W_{\psi}(y,\mu)$ is not fixed
uniquely by the above properties, and the valley is an
approximate classical solution that interpolates smoothly
between a soliton-antisoliton pair at large separation
and the trivial vacuum.
\pr

A
general method for constructing
valleys in two-dimensional models has been suggested in
ref. \cite{dorinst}.
It identifies the valley parameter $\mu$ with time, and
considers the valley as a kink-antikink scattering problem. The
reliability of this identification was conjectured in ref.
\cite{dorinst} to be a general property of theories whose instantons
become solitons in one dimension higher. Following ref. \cite{dorinst}
let $\psi_*$ denote a soliton-antisoliton scattering solution,
obeying by construction the time-dependent equation
\be
-\frac{\partial ^2 \psi _{ *}}{\partial \mu ^2}=
\frac{\delta S_0 [\psi _{ *}]}{\delta \psi_{*}}
\label{tessepta}
\ee
We see immediately that if we choose as weight function
\be
W_{\psi}(y,\mu)=-\frac{\partial ^2 \psi _{*}}{\partial \mu ^2}/
\frac{\partial \psi _{*}}{\partial \mu}
\label{unnumberred}
\ee
equation (\ref{weight}) becomes
identical to equation (\ref{tessepta}). In
the case of
the $O(3)$ $\sigma$-model, the weight function (\ref{unnumberred})
is positive
definite and decays exponentially as $|y|$ tends to infinity. The
boundary condition (\ref{tessexi}) is
automatically satisfied by the
scattering solution $\psi_*$, which is therefore an exact valley
solution of the $O(3)$ theory. The introduction of the conformal
symmetry breaking
term (\ref{tessdio}) complicates the problem, but we follow
ref. \cite{khoze} in assuming that it does not change significantly
the valley configuration, but only has the effect of introducing
an infrared cutoff in the valley action when either the instanton
or the anti-instanton becomes large. In the case of $O(3)$ $\sigma$-model
for $2\pi m << g^2$
the effect of this term on the valley path-integral
amounts \cite{BW}
to introducing a harmless
pre-factor $e^{-S_1}$
of order one.
\pr

\nk{\it Reduction of the $SL(2,R)/U(1)$ coset model}
\pr
The above procedure can be applied to the $SL(2,R)/U(1)$ model
\cite{witt} of interest to us,
but with certain differences which we shall
mention as we proceed. We start from the action (\ref{Yung 11}),
concentrate
on radially-symmetric configurations $w=f(r)e^{-iq\theta}$, and
restrict our attention to the one-instanton sector. Setting
$y=lnr$, the action (\ref{Yung 11}) becomes
\be
S_{red} =\int _{-\infty}^{+\infty} dy \frac{1}{1+f^2}[(f '   )^2
+f^2]
\label{tessokto}
\ee
and redefining $\phi=c+sinh^{-1}f$ we get
\be
S_{red}=\int _{-\infty}^{+\infty} dy [(\phi ')^2
+ 1 - \frac{1}{cosh^2\phi}]
\label{tessennea}
\ee
The corresponding equations
of motion remain unchanged if the constant
is
subtracted from  (\ref{tessennea}).
Then,
the effective
action of the model becomes
\be
S_{red}=\int _{-\infty}^{+\infty} dy [(\phi ') ^2
- \frac{1}{cosh^2\phi}]
\label{tessdeka}
\ee
There are static solitons which are extrema of the action
(\ref{tessdeka}) of the form
\be
\phi (y,y_0)=\pm sinh^{-1}(\sqrt{2}(y-y_0)
\label{tessenteka}
\ee
where the prefactors $+$ ($-$) correspond to kink (antikink)
solutions respectively, which are sketched in figure 4. It is
easy to check that these solitons have finite energy, like the
sphalerons:
\be
E=\int _{-\infty}^{+\infty} dy [(\phi ')^2 + |V(\phi)|^2]
= \frac{3\pi}{\sqrt{2}}
\label{tessedodeka}
\ee
It is interesting to notice that
these solutions, when expressed in terms of
$\sigma$-model variables,
correspond to
monopoles on the world-sheet, with unit charge
\cite{ovrut}, centred at the origin of the
stereographically-projected plane, as
becomes evident from the discussion in
section 2.
In the string case, monopoles
have been interpreted \cite{emndua}
as the singularities of black holes in target space,
with the charge $q_m $ playing the r\^ ole of the
black hole mass in units
of the Planck mass, whilst the corresponding
antimonopoles required by the zero net ``spikiness''
of a compact world-sheet represent their horizons.
Within
this interpretation the kink-antikink
solutions should be viewed as
microscopic (Planck mass)
black holes in
the space-time foam.
\pr
Having identified the black holes with kink solutions in the
reduced model (\ref{tessdeka}), a
natural question concerns the r\^ ole of
two-dimensional world-sheet instantons, which, as we have argued,
describe massive string mode effects, and induce
transitions between black holes of different mass. To answer this
question it is necessary to consider time-dependent solutions
of (\ref{tessdeka}) by solving the equation
\be
S(\mu, y) \equiv
\frac{\partial ^2 \phi}{\partial \mu ^2} -
\frac{\partial ^2 \phi}{\partial y^2} - \frac{1}{2}
\frac{\partial |V(\phi)|}
{\partial \phi}=0
\label{tessdekatria}
\ee
where $V(\phi)=-1/{cosh^2{\phi}}$, $\mu$ is a Minkowskian time
variable, and $y$ is a space variable.
The full time-dependent equation for
$F=sinh{\phi(\mu,y)}$ reads
\be
\frac{F}{1+F^2}[-(\frac{\partial F}{\partial \mu})^2
+ (\frac{\partial F}{\partial y})^2 +1]=-\frac{\partial ^2 F}
{\partial \mu^2} + \frac{\partial ^2 F}{\partial y^2}
\label{tessdekexi}
\ee
whose solutions can be split into two branches:
\bea
&~&- \frac{\partial ^2 F}{\partial \mu^2}    +
 \frac{\partial ^2 F}{\partial y^2}    =0   \nn \\
&~&-(\frac{\partial F}{\partial \mu  })^2
+(\frac{\partial F}{\partial y})^2 + 1=0
\label{tessdekeptaa}
\eea
and:
\bea
&~&- \frac{\partial ^2 F}{\partial \mu^2}    +
 \frac{\partial ^2 F}{\partial y^2}    =F   \nn \\
&~&-(\frac{\partial F}{\partial \mu  })^2
+(\frac{\partial F}{\partial y})^2 =F^2
\label{tessdekeptab}
\eea
The branch of solutions ({\ref{tessdekeptaa}) admits
Lorentz-boosted soliton solutions
\be
\phi(y,\mu)=\pm sinh^{-1}[\frac{y-y_0}{\sqrt{1-u^2}}+
\frac{u\mu}{\sqrt{1-u^2}}]
\label{tessdekatess}
\ee
where $u$ is the velocity of the soliton
smaller than unity.

The other branch (\ref{tessdekeptab}) resembles
a massive Klein-Gordon equation and the
corresponding condition for vanishing of the action.
Solutions of this branch
assume the form
\be
  F(\mu, y) =e^{\pm \frac{u \mu}{ \sqrt{1-u^2}}\pm\frac{y-y_0}
 {\sqrt{1-u^2}}}
\label{inst}
\ee

\nk Upon passing into the $\sigma$-model formalism,
such solutions describe instantons, centered at the origin
of the world-sheet, with scale sizes determined
appropriately by $\mu$ and $y_0$.
Notice that
the static ($\mu$-independent ) configurations (\ref{inst}), are
not finite -energy solutions
of the action (\ref{tessdeka}), but have a logarithmic infrared
singularity analogous to that of the topological charge (\ref{Yung 21}).
\pr
\nk{\it Monopole-Antimonopole Valley in the
reduced $SL(2,R)/U(1)$ model}
\pr
We are now ready to demonstrate the existence of
monopole-antimonopole and instanton-anti-instanton
valleys in the
$SL(2,R)/U(1)$ model.
To this end, following the conjecture of ref. \cite{dorinst},
we shall consider scattering solutions of the time-dependent
equation (\ref{tessdekatria}), finding
approximate
solutions to the scattering of a kink and an
antikink. To understand physically what is going on, we recall that
a valley
is a field configuration that interpolates smoothly between
the trivial vacuum and a far separated soliton-antisoliton
pair. In the prescription of \cite{dorinst}, which we follow here,
the `time' $\mu$
in equation (\ref{tessdekatria}) corresponds
to the separation of the soliton-anti-soliton, and thus to
the valley separation parameter. In a conformally
invariant theory, the concept of a finite-separation valley
is meaningless. Thus
conformal symmetry breaking terms must
be introduced.
The physics of the valley configuration is supposed to be
independent
of the particular form of the symmetry breaking term.
\pr
In the case of the $SL(2,R)/U(1)$
model
there are two kinds of such symmetry breaking terms.
One corresponds to classical symmetry-breaking :
light-string-matter background deformations of the $\sigma$-model,
which are by themselves relevant perturbations, causing a flow
of the system towards a non trivial fixed point, and thus
leading to the interpretation of the
renormalization scale
parameter as target time\cite{emnqm}, as
discussed in the next section.
The form of such deformations
is found by $W_{\infty}$-symmetry
considerations in the target space
of the $SL(2,R)/U(1)$ model; in fact one constructs marginal
deformations out of $W_{\infty}$ currents, which assume the form
(\ref{margintax}):
\be
L_0^1{\overline L}_0^1 \propto
\Phi ^{c-c}_{\frac{1}{2},0,0} + i(\psi^{++}-\psi^{--})) +\dots
\label{margintach}
\ee

\nk where the massive string modes $i(\psi ^{++}-\psi^{--})+ \dots$
are given in terms of $SL(2,R)$ currents \cite{chaudh}.
The operator
$\Phi ^{c-c}_{\frac{1}{2},0,0}(r)$
is given in (\ref{tachyon}),
and generates
the light string matter.
As discussed in sec. 2,
the  deformation
$\Phi ^{c-c}_{\frac{1}{2},0,0}(r)$
alone is relevant, leading
to a breaking of the conformal invariance of the model.
The truncation of the marginal deformation (\ref{margintach})
to its relevant light-matter part has physical significance,
reflecting the fact
that local scattering experiments cannot observe the
delocalized (solitonic) massive string modes.
\pr
When
correlation functions of such light-matter
deformations are
evaluated in the presence of
monopole-antimonopole configurations, the corresponding valleys
in the $\sigma$-model theory
correspond to saddle-point
configurations in the path integrals for these correlators, leading in
turn to non-factorizable
contributions to the $\nd{S}$-matrix for the
light string modes.
However, in such computations,
the deformations appear as scattering terms in a
theory whose action is conformally invariant.
One
needs additional conformal symmetry breaking terms to
give meaning to finite-separation valleys.
These are provided by dilaton terms, which constitute the second
kind of conformal symmetry breaking terms.
They appear at the quantum level and assume the form
\be
S_{dil}=-\int d^2z \frac{1}{8\pi}
\sqrt{h} R^{(2)} ln(1 + |w|^2) + \dots
\label{dilaton}
\ee
This term does not have any coefficient $k$, unlike the
Wess-Zumino action term (\ref{Yung 11}). Redefinition of the fields
$r$ by a
factor $\sqrt{k}$, so as to ensure a canonical kinetic term,
leads to a rescaling of
the conformal symmetry breaking term by a
factor $\frac{1}{\sqrt{k}}$, reflecting
its quantum nature\footnote{We
remind the reader that in the stringy Wess-Zumino models (\ref{Yung 11})
the level parameter $k$ is inversely proportional to the string
Regge slope $\alpha '$. The later is a quantum-effect book-keeping
parameter.}. For relatively large $r$, such terms
become
\be
S_{dil}=-\frac{1}{8\pi\sqrt{k}}\int d^2z \sqrt{h} R^{(2)} r(z,{\bar z})
\label{larger}
\ee
Concentrating all the curvature at a single point $z^*$ we have,
for surfaces with the topology of a sphere  (Euler characteristic
$\chi=2$), $R^{(2)}=8\pi \delta^{(2)}(z-z^*)$, and thus
\be
 S_{dil} =-\frac{1}{\sqrt{k}} r(z^*)
\label{approxdil}
\ee
Hence, in the large $k$ limit,
the dilaton term becomes important
for target space points at a distance $r >> 1$, far
from the
black-hole
singularity at $r=0$. For finite $k$ (as is the case of critical
strings $c=26$), the dilaton term becomes important in a region
localized around
the singularity $r=0$.
\pr
The scattering solution may be found by first noticing
that the function
\be
{\overline \phi}(y, \mu)=sinh^{-1}(\mu/cosh(y))
\label{scatter}
\ee
is an exact
solution of the equation (\ref{tessdekatria}),
with a potential
modified by a dilaton source-term (\ref{dilaton}).
We can
demonstrate this
analytically by computing the
the left-hand-side $S(\mu,y)$
of eq. ({\ref{tessdekatria})
for the function (\ref{scatter}). The result is
\bea
S(\mu,y)&=&\frac{y}{cosh^2\mu \sqrt{\mu^2 + cosh^2 y}} =
\frac{1}{cosh^2 y} tanh[\phi (\mu,y)] = \nn \\
=\frac{1}{cosh^2 y} ~&\times&|\frac{\delta}{\delta \phi (\mu,y)}
ln(cosh(\phi (\mu, y))|_{\phi = {\overline \phi}}
\label{potentialaddd}
\eea
Thus, the function (\ref{scatter}) is an {\it exact} solution
of a $\sigma$-model whose potential energy has been modified
by a conformal-symmetry breaking term of the form
\be
-2\int d\mu dy \frac{1}{cosh^2y}   ln[ cosh\phi (\mu, y) ]
\label{potadd}
\ee
{}From figure 5, it is clear that for large $\mu$
the valley solution (\ref{scatter}) yields a
far-separated pair of a boosted
kink and an antikink approaching each other
with an ultra-relativistic velocity $u \simeq 1 $.
\pr
When expressed in terms of the original $\sigma$-model
variables this relation yields :
\be
 -2\int d\mu |z| d|z| \frac{1}{(|z|^2 + 1)^2} ln(1 +
 |w(z, {\bar z};\mu)|^2)
\label{sigmavar}
\ee
One should now remember that we are effectively working
on a world-sheet which has the topology of a sphere, and
is projected stereographically onto the complex plane.
Without loss of generality we can fix the geometry to
be that of a sphere of
radius $1/2$, and, hence, of curvature $8$. In that case
there is an induced
$O(3)$
invariant measure on the complex plane
defined by \cite{ovrut}
\be
d^2 z  \sqrt{|g(z,{\bar z})|}=\frac{1}{2(1 + |z|^2 )^2} d^2z
\label{othreemes}
\ee
thereby implying that (up to an overall
normalization) the
conformal symmetry breaking term (\ref{potadd})
coincides with the
the usual $\sigma$-model dilaton term (\ref{dilaton})
after a homotopic extension in $\mu$.
The normalization factor in this case determines the
value of $k$ corresponding to the solution (\ref{scatter}), and
a simple calculation yields  $k=\frac{1}{4}$.
To understand this value, one should note that
(\ref{dilaton}) is only correct in the large $k$
approximation. If instead one assumes that the
theory (\ref{Yung 8}) together with (\ref{dilaton})
represents an exact conformal field theory for all $k$,
then the corresponding central charge can be computed
from the asymptotically linear dilaton term (\ref{dilaton})
far away from the singularity. The corresponding central
charge is \cite{witt}
\be
c=2 + 6/k
\label{centraltwo}
\ee
for large $k$, which
agrees with the exact
answer (\ref{central}) within an error $1/k^2$.
For $k=1/4$ eq. (\ref{centraltwo})
yields $c=26$, i.e. the target-space interpretation remains.
However, there are, of course,
corrections
in higher orders in $\frac{1}{k}$ in the expression
for the dilaton and graviton which we do not discuss here.
We assume that the qualitative features
of the valley remain the same for the exact black hole solution
$k=9/4$. As we shall see later, the results of importance for
us are, in any case, independent of the details of the
valley configuration.
\pr
To check whether (\ref{scatter}) is a sensible valley configuration,
we should verify that the corresponding weight function
(\ref{unnumberred}) is positive definite (for $\mu \ge 0$), and
decays fast enough for large $y$.
{}From (\ref{scatter}),(\ref{unnumberred}) we then have
\be
W(\mu, y)= \frac{\mu sech^2(y)}{1 + \mu^2 sech^2(y)}
\label{vall}
\ee
which is indeed positive definite and decays
exponentially with large $y$. We also note as a curiosity
that
the valley function (\ref{vall})
coincides formally with the corresponding
function for the $O(3)$ $\sigma$-model \cite{dorinst}.
It can easily be checked that, for highly separated
monopole-antimonopole pairs, the solution
(\ref{scatter}), when substituted back to the
Wess-Zumino action, yields the characteristic
logarithmic infinities of an
isolated monopole and an anti-monopole \cite{ovrut}.
Also, for zero separation $\mu \rightarrow 0$, it
induces the collapse into the trivial vacuum, since the corresponding
action vanishes.
When mapped back to the original
$\sigma$-model variables, upon making the
replacement $\mu \rightarrow v -\frac{1}{v}$, with $v$ the
conventional
$\sigma$-model separation parameter,
the valley configuration
yields a concentric valley,
\be
 w(z,{\bar z})=\frac{(v-1/v){\bar z}}{1 + |z|^2}
\label{cocentric}
\ee
The latter may then
be mapped to the
desired valley by applying a conformal transformation
in the world-sheet, involving an appropriate inversion followed
by a translation, in a similar spirit to the case of $O(3)$
$\sigma$-model \cite{dorinst}.
The function (\ref{cocentric}) is shown in figure 6.
\pr

\nk{\it Instanton-Anti-instanton Valley in the
reduced $SL(2,R)/U(1)$ model}
\pr
The method of the previous subsection for the
construction of the $\sigma$-model
valley using
the
soliton-anti-soliton
scattering
solution of
the
homotopically-extended (`time-dependent') reduced model
(\ref{tessdeka}), cannot be applied directly to the
case of valley configurations for instanton-anti-instanton
pairs in the reduced model of the form (\ref{inst}).
The latter
are delocalized classical solutions
of (\ref{tessdeka}), that do not have finite energy.
This complicates technically the construction
of a classical `scattering' solution, because
the definition of an asymptotic (freely-propagating)
state in the
infinite future needs elaboration.
\pr
However, this does not imply that a valley configuration
cannot be constructed. Recalling that the latter
is a smooth interpolating function between
an infinitely-separated pair of topological defects and the
trivial vacuum, we observe that
all
one needs for the construction of a valley
in the reduced-model formalism  is the
first half of a `scattering-like' process, i.e., the part
that extends
from the
infinite past, where the defects are infinitely separated,
till the time their centres come
close to each other.
This last stage corresponds to the collapse of the valley into the
trivial vacuum.
Thus, as a trial valley for our case,
we consider,
following \cite{khoze,BW},
a simple superposition
of a instanton and an anti-instanton (\ref{inst}),
\be
g(y,\mu)=sinh^{-1}[exp(y+\mu)]-sinh^{-1}[exp(y-\mu)]
\label{suminst}
\ee
We observe that (\ref{suminst}) can be derived from
a monopole-antimonopole pair
\be
sinh^{-1}(y+\mu)-sinh^{-1}(y-\mu)
\label{monopolpair}
\ee
by
a conformal transformation in the $(y,\mu)$ plane :
\bea
\mu + y \rightarrow e^{y + \mu} \nn \\
y - \mu \rightarrow e^{y - \mu}
\label{confo}
\eea
To be more accurate, recall that
the scattering solution (\ref{scatter})
represents to a very good approximation a monopole-antimonopole
pair approaching each other at an ultra-relativistic
velocity $u \simeq 1$. This implies that the pertinent
conformal
transformation that connects (\ref{suminst}) to the
monopole-antimonopole valley solution (\ref{scatter})
is
\bea
\mu + y \rightarrow e^{\frac{y + \mu}{\sqrt{1 -u^2}}}
\nn \\
y - \mu \rightarrow e^{\frac{y - \mu}{\sqrt{1-u^2}}}
\label{conf2}
\eea
It should be stressed
that the transformation (\ref{conf2}) is only {\it approximate}.
Due to the velocity $u$, the actual transformation that connects
the
instanton valley with the monopole-antimonopole valley is
{\it not  conformal}, since it involves Lorentz-boosted
solutions. However, for $u \simeq 1$ such non-conformal
effects
may be ignored, and hence one is effectively
working with light-cone coordinates $y \pm \mu$,
which can be complexified in the
Euclidean formalism to yield
the
usual $z, {\bar z}$ variables of the complex $(y,\mu )$
plane (caution: this is not the original complex plane of the
Wess-Zumino model).
\pr
One may  express the time-dependent action that reproduces
(\ref{tessdekatria})
in light-cone coordinates, and then apply the transformation
(\ref{conf2}). The kinetic term is conformally invariant,
whilst the potential term changes as follows:
\be
(1-u^2)\int \frac{dz'd{\bar z}'}
{|z'|^2} \frac{1}{cosh^2\phi'(z',{\bar z'})}
\label{pottrnsf}
\ee
where we took into account the fact
that the field $\phi$ is a scalar under
coordinate transformations. Due to the $(1-u^2)$ factor,
the potential
term becomes important only for small $|z|\simeq O(1-u^2)$,
where it has a
similar form to the original potential term, yielding
(\ref{tessdekatria}). This is the region where the original potential
term is dominant for intermediate and/or large separations $y$.
Thus,
we can argue that the
instanton valley may be derived,
to a good approximation,
by applying
the transformation
(\ref{conf2}) to the exact monopole-antimonopole valley.
Taking into account the above remarks on the
approximate invariance of the valley
action under conformal
transformations,
we observe that the weight function for the
instanton valley is induced by the
conformal transform of the monopole weight function (\ref{vall}),
thereby maintaining all the necessary properties.
We therefore consider the following representation of
the instanton
valley, expressed in terms of the original $\sigma$-model variables
\be
w(z,{\bar z})_{inst} \simeq \frac{{\bar z}^{\frac{1}{\sqrt{1-u^2}}}
sinh(\frac{\mu}{\sqrt{1-u^2}})}{cosh(|z|^{\frac{1}{\sqrt{1-u^2}}}
cosh[\frac{\mu}{\sqrt{1-u^2}}])}
\label{confval}
\ee
The function (\ref{confval}) is represented in figure 7.
That this
function corresponds to a valley is guaranteed by the
general property that a conformal transformation of a valley
is also a valley configuration \cite{BW,dorinst}.
As in the monopole-antimonopole valley case, the
conformal symmetry breaking term in the action that
is necessary to give a meaning to the finite-separation
valleys is provided by the dilaton term (\ref{dilaton}).
\pr
\nk {\it Contributions to Correlation Functions}
\pr
Having constructed approximate valley configurations
in the $SL(2,R)/U(1)$ model, we now discuss
the physical consequences for the
correlation functions of the matter operators (\ref{tachyon})
$\Phi_{\frac{1}{2}, 0,0}^{c-c} (r)$. From a world-sheet
point of view, such correlators are nothing but complicated
algebraic functions of $\sigma$-model fields $r(z,{\bar z })$.
The above-described valley configurations are, then, viewed
as dominant configurations in a path-integral
evaluation of the forward scattering amplitude
on the world-sheet, which by means of the optical theorem
is related to the total cross section for the fields
$r(z, {\bar z})$. For completeness, we
mention that
the
conformal symmetry breaking terms in this picture
allow for non-vanishing inelastic scattering amplitudes
on the world-sheet.
\pr
Although from a world-sheet point of view one can define
perfect quantum-mechanical scattering in the presence of
valley configurations, and apply the optical theorem
as a consequence of world-sheet
unitarity,
this is not the case in target space. As a result of the
valleys,
there are
obstructions
to
the
interpretation of the tachyon correlators
as generating functionals for target-space scattering
amplitudes. In the case of black-hole $SL(2,R)/U(1)$
string models,
the
spherically-symmetric four-dimensional
target space is obtained \cite{emndua}
as a
topologically non-trivial
homomorphism of the world-sheet onto a two-dimensional
target subspace. Under such
a mapping, the
non-separable contributions of the
valleys
to the world-sheet scattering matrix result - as we shall
discuss in sec. 6 -  in
non-analytic contributions to the target-space $S$-matrix,
thereby leading to the $\nd{S}$-formalism.
This result
has been anticipated in \cite{emnqm} using a pure
$\beta$-function approach.
Here we
rederive it
using
an
alternative - and physically appealing -
point of view
exploiting the topologically non-trivial structure
of the
world-sheet of a string
propagating in target-space
black-hole backgrounds.

\section{Renormalization-Group Scale as
Target Time}
\pr
We review in this section our interpretation
of the renormalization group scale in the subsystem
of the light string modes as target time. Most of the
material is contained in ref. \cite{emnqm}, and here we give
only the general idea and technical details that are needed
for the development of the material presented in this
article.
\pr
We use the concept of the local (on the world-sheet)
renormalization group equation, which was originally
introduced as a formal tool for analyzing $\sigma$-models
\cite{shore}. It was used to prove the ``off-shell'' corollary
of the c-theorem \cite{osborn}, according to which
perturbative $\beta$-functions are gradient flows
of the string effective action in target space,
but no physical significance was attached to the
local dependence of the cutoff. However, we identify the local
cutoff as the Liouville mode, whose kinetic term has a
temporal signature for supercritical strings ($c > 26$) \cite{aben}.
This was the first paper where cosmological time
was introduced into string theory by
exploiting the
flow of the subsystem pertaining to the extra compact
dimensions under the renormalization group. Subsequently,
this formalism was interpreted in terms of Liouville theory
\cite{polch}, but there was not any serious attempt, at the time,
to associate the target time with a local renormalization
group scale on the world-sheet in the usual $\sigma$-model sense,
 and exploit the consequences of such a formalism.
As we have already pointed out, the inclusion of topological
fluctuations (such as monopole-antimonopole or
instanton-anti-instanton pairs) in a critical string
theory (such as the $SL(2,R)/U(1)$ coset model with
$k=9/4$) makes it supercritical, necessitating the introduction
of such a time-like Liouville field, whose interpretation
as target time we now explain.
\pr
Consider a general fixed-point world-sheet theory, such as
the critical black hole model\footnote{The fact that this
classical background configuration depends only on target space is
already suggestive of a {\it time} interpretation of the
renormalization group cutoff.}. Upon deforming the theory
by light matter background fields, one gets formally for the
$\sigma$-model action
\be
S_{def}=\int d^2x (L_0 + \Phi (x)R^{(2)}(x) - \Sigma  (x) +
g^i(x) V_i )
\label{actsigm}
\ee

\nk where the renormalized couplings
$g^i(x)$ depend on the world-sheet coordinates $x$ through
the local renormalization group scale $-ln a(z,{\bar z})$
to be identified
as the Liouville mode $\phi (x)$. The corresponding
normal-ordered vertex operators $V_i(x)$ are also
dressed by the Liouville field cut-off, so as to be
$(1,1)$ operators at the fixed point. Simple power
counting arguments show that the extra `dilaton' and
`tachyon' counterterms $\Phi (x)$, $\Sigma (x)$ in
(\ref{actsigm}) are necessary when renormalizing in
curved space-time, although they vanish
in any global renormalization group scheme. The
explicit form of the `tachyon' counterterm in, for example,
dimensional regularization is
\be
\Sigma _{Bare} = a^{\epsilon} [\Sigma  + L_{\Sigma }(g^i)]
\label{counteterms}
\ee

\nk where $a$ is the ultraviolet cut-off introduced
in section 2,$\epsilon$ is an anomalous dimension, and
\be
   L_{\Sigma }=\frac{1}{2}{\cal G}_{ij}\partial _{\alpha}g^i
\partial ^{\alpha} g^j
\label{tachyoncount}
\ee

\nk and, as already mentioned,the
$g^i$ depend on the local
renormalization group scale (Liouville mode) $\phi(x)$.
It can be shown \cite{osborn} that ${\cal G}_{ij}$ is
related to divergences of the two-point function
$<V_iV_j>$, but it differs from the Zamolodchikov
metric, so that the positivity for unitary theories is
not immediate: however, this does not concern us here.
The $\beta$-functions corresponding to $\Sigma  (x)$, $g^i$
are defined by
\bea
{\hat \beta}_{\Sigma } \equiv -d\Sigma  /dln(a) =
\epsilon \Sigma  + \beta _{\Sigma }(g)  \nn \\
{\hat \beta}_{i}
\equiv -dg^i/dln(a) = \epsilon g^i + \beta^i (g)
\label{beta}
\eea

\nk where
\be
\beta_{\Sigma }=
\frac{1}{2}\chi _{ij}\partial _{\alpha}
g^i \partial^{\alpha} g^j =\frac{1}{2}\beta ^i \chi _{ij} \beta ^j
\partial _{\alpha} \phi \partial _{j} \phi
\label{chi}
\ee

\nk The renormalizability of the model
(\ref{actsigm}) now specifies $\chi _{ij}$  in terms
of ${\cal G}_{ij}$. Imposing the invariance of (\ref{actsigm})
under the renormalization group operator
\be
(\epsilon - {\hat \beta}^i\partial _i - {\hat \beta}^\lambda
\partial_{\lambda}) S = 0 \qquad : \qquad \lambda \equiv
(\Phi, \Sigma )
\label{rgo}
\ee
we find
\be
\chi _{ij} =(\epsilon - {\hat \beta}^k\partial _k )
{\cal G}_{ij}-(\partial _i {\hat \beta}^k {\cal G}_{kj}-
(i \leftarrow\rightarrow j) )
\label{chigi}
\ee

\nk In the limit $\epsilon \rightarrow 0$, the `tachyon' $\Sigma (x)$
contributes the following terms to the $\sigma$-model action :
\be
-{\cal G}_{ij}^{(1)} \beta^j(g^i + \beta^i\phi)
\partial _{\alpha}
\phi \partial ^{\alpha}\phi \equiv
G_{00}\partial _{\alpha}
X^0\partial ^{\alpha}X^0
\label{backmetr}
\ee

\nk where ${\cal G}_{ij}^{(1)}$ is the residue of
the
simple
$\epsilon$-pole\footnote{In covariant cut-off
approaches
this should
be replaced by the leading logarithmic divergences, but
the qualitative results remain the same.}. We now
see that
$G_{00}$ may be viewed as a
perturbation in
the temporal component
of the background metric in the two-dimensional black-hole
string model, and denote the Liouville mode $\phi (x)$
henceforward as the target time $X^0 (x)$.
Notice that
in this picture
the Liouville field
acquires non-trivial corrections to its kinetic term
as compared to eq. (\ref{liouv}).
We see
clearly that the temporal (cutoff) dependence of the metric
tensor is a back reaction due to matter emission.
Indeed, if the matter couplings $g^i$ were marginal,
then the $\beta^i$ would vanish, as would the corrections
to the background metric.
\pr
At this stage it is useful to compare the above
picture to that of ref. \cite{aben}. Contrary to
that reference, in our approach
the target time
(local renormalization
group scale) appears already in the fixed-point
action (\ref{Yung 8}). At first sight this seems
strange from a renormalization group point of view.
Indeed, even in the case of a local scale $\phi (z,{\bar z})$,
the corresponding
renormalization group equation for the effective action  $Z$
reads\cite{shore}:
\be
[\frac{\partial}{\partial \phi (z,{\bar z})}
+ \beta^i(g^i(\phi))
\frac{\partial}{\partial g^i(\phi)} ]Z=0
\label{localren}
\ee
In our interpretation
of target time as a
renormalization group scale, this equation should be
considered as a constraint on the physical states.
In the Liouville theory framework, which we follow here,
the constraint (\ref{localren})
is incorporated automatically as a result of the
conformal dressing of the various $\sigma$-model
operators \cite{david}. In the conformal
Wess-Zumino theory, the constraint (\ref{localren}) is satisfied
as a result of the world-sheet equations of motion of the
$\theta$ field, which
in our language is the Liouville mode/time
\footnote{The precise
functional relation between $\theta$ and $X^0$ \cite{Moore,kog}
is not directly relevant for our purpose here.} $X^0$.
This is the crucial formal difference of our approach
from that of ref. \cite{aben,david}.
The Liouville field at the fixed point
must
satisfy the equation of motion of a free $\sigma$-model
field. Under this condition,
any apparent time-dependence in the action (\ref{Yung 8})
disappears at the level of the background fields, which are static,
and hence their local renormalization group $\beta$-functions
vanish. It should be stressed at this point, that these
$\beta$-functions are {\it not} the $\beta$-functions
leading to the black-hole solution of \cite{witt}.
{}From that point of view the Liouville mode
is the spatial coordinate, because in that picture
the string is subcritical with
$c=1$ \cite{witt}.
\pr
The arrow of target time can now be found by analyzing
the correlation functions $A_N =<V_{i_1} \dots V_{i_{N}}>$
of matter vertex operators $V_i$. We use two
representations of the same physics. One describes the Euclidean
black hole without matter deformations as a subcritical
$c=1$ string model on a flat target-space-time with a modified
cosmological constant term \cite{kutasov}
\be
2\pi \nu \beta {\bar \beta} exp(-\frac{2}{\alpha _{+}}\zeta)
\label{area}
\ee

\nk Here,
$\beta$  and $ {\bar \beta}$ are  chiral ghost fields
and $\zeta$ is a free field
entering the representation
of the $SL(2,R)/U(1)$ current algebra, $\nu$ is
the black-hole mass $M_{bh}$,
and
$\alpha _{+}=2k-4$
is related to the level parameter $k$
by the requirement
that the conformal dimension of the modified
area term (\ref{area}) be unity.
Because the matter field describing
target time is compact in this picture,
it represents a Euclidean black hole. The second
representation, applicable to a Minkowski black hole, is to
use marginal deformations of the coset model that respect
the $W_{\infty}$ coherence-preserving symmetry.
As mentioned earlier, truncating the theory
to the light propagating modes corresponds
to including the instantons that represent
massive modes as discussed in section 2,
which make the conformal
theory supercritical with $c > 26$, resulting
in a time-like local cutoff. This latter
representation is the more relevant for our purpose here,
but it is instructive to compare the two formalisms.
\pr
In the $c=1$ Euclidean picture vertex operators
for the simplest $(1,1)$ $SL(2,R)$ primary states
take the form
\be
V_{j,m}^T \propto \gamma^{j-m}exp(\frac{2j}{\alpha _{+}}\zeta)
exp(im\sqrt{2/k}X)
\label{primary}
\ee

\nk where the boson $X$ is compact
with radius $\sqrt{k}$ and represents Euclidean time, as
mentioned above,
$\gamma$ is
another chiral boson
in the
free-field realization of the
$SL(2,R)$ current algebra, and
$(j,m )$ is the
$SL(2,R)$ isospin
and its third component, which are subject to the on-shell
condition
\be
-\frac{j(j+1)}{k-2}+\frac{m^2}{k}=0
\label{onshell}
\ee
The $W_{\infty}$ symmetry algebra tells us that the deformation
(\ref{primary}) is exactly marginal only in the infinite-mass limit
for the
heavy string modes. The integration over the zero mode of $\zeta$
in the path integral for a correlation function $A_N$
may be performed by inserting the formal identity
\be
\int dA \delta (\int d^2z |\beta |^2 e^{-\frac{2}{\alpha _+}\zeta}
-A) =1
\label{constraint}
\ee

\nk resulting in the representation
\be
A_N=\nu ^s\Gamma (-s) \times \langle V_{j_{1}m_{1}}\dots
V_{j_{N}m_{N}}[\int d^2z |\beta|^2 exp(-\frac{2}{\alpha ^+}\zeta (z)]^s)
\rangle _{\nu=0}
\label{correl}
\ee

\nk where
\be
\Gamma (X) =\int _0^\infty dA A^{-s-1}e^{-A}  \qquad ; \qquad
 s=\sum _{i=1}^{N}j_{i} +1
\label{gamma}
\ee
and $\nu$ is the
coupling constant of the modified
area term (\ref{area}) which is related to
the black-hole mass.
The symbol $<\dots >_{\nu=0}$ denotes the
integration over $X$ and the world-sheet dependent
non-zero modes of $\zeta$, in the limit of
vanishing mass for the black hole.
\pr
Because of the $\Gamma (-s)$ factor in (\ref{correl}),
the unregularized correlation functions $A_N$
are ill-defined whenever a discrete massive string state
is excited. In this case, $s$ is a positive integer $n^+$ and
one can regularize the $\Gamma (-n^+)$ pole by analytic
continuation, replacing the integral representation
(\ref{gamma}) of $\Gamma (-s)$ by \cite{kogan2}
\be
I(s)_{reg}\equiv \int _C A^{-s-1}exp(-A)dA
\label{regul}
\ee

\nk where the contour $C$ is shown in figure 8.
This introduces imaginary parts in the correlation
functions :
\be
I(n^+)=lim_{\epsilon \rightarrow 0}[exp(-2\pi i\epsilon) -1]
\Gamma(-n^+-\epsilon) =(-1)^{n^{+}}\frac{2\pi i}{(n^+) !}
\label{imag}
\ee

\nk which can be interpreted as instabilities
reflecting the renormalization group flow of the theory.
The parameter $A$ appearing in (\ref{regul})
is naturally interpreted as the effective area
of the world-sheet as measured in target space. This
follows from the way
in which
(\ref{constraint})
is related to the free field $\zeta$ and the ghost
fields $\beta$ and ${\bar \beta}$, which is the same as in the
cosmological constant term (\ref{area}) in the world-sheet
action. However, although the $\phi$ field of the
subcritical $c=1$ model is analogous to the Liouville mode
discussed earlier, it is space-like rather than time-like,
and we need to extend the above discussion to the
supercritical Minkowski black hole representation.
\pr
In this case, as already discussed,
a time-like Liouville field can be introduced
to restore criticality in the sense that the total central charge
$c=26$, and one finds a `tachyon' area-like Liouville
term similar to (\ref{area}), with $\nu$ having the
traditional interpretation as a cosmological
constant on the world-sheet. The standard analysis
\cite{david}
of Liouville theory correlation functions then applies, leading
again to expressions of the form (\ref{correl}), where the
parameter $A$ in the representation of
the $\Gamma$-function regularized \`a la (\ref{regul})
now admits a world-sheet area interpretation. The
imaginary parts (\ref{imag}) of the correlation functions
can now be interpreted directly as instabilities of the
string vacuum. In the supercritical string, deformed by light matter
and truncated to its light subsystem, this is just the usual
flow of the latter from one fixed point towards another with
lower central charge $c$, in accordance with
Zamolodchikov's c-theorem \cite{zam} for any unitary theory.
\pr
It is evident from the form of the contour integral (\ref{regul})
illustrated in figure 8 that
there are two phases in the renormalization
group flow. It starts from an infrared fixed point with large
world-sheet area, corresponding to a large negative value
of the Liouville field, passes through an ultraviolet
fixed point with small world-sheet area, corresponding to
a large positive value of the Liouville field, and
returns to the infrared fixed point. In this picture,
it is natural to regard the ultraviolet fixed point
as the end-point of the actual time flow in target space-time.
This interpretation is suggested by the identification
made in the previous paragraph of the renormalization
group flow with the decay of a metastable vacuum\footnote{This
is an explicit demonstration of a generic conjecture made in
ref. \cite{kogan2}.}. Our interpretation is reinforced by
the analogy with the divergences in the path integral
representation
of `bounce' solutions in ordinary field theory \cite{bounce}.
For the convenience of the reader we give here the formula for
the decay probability per unit volume $\Gamma /V$
of a false vacuum state $\phi _f$
in the conventional field theory case
\bea
\Gamma/V \propto \frac{B^2}{ \hbar }exp(-\frac{B}{\hbar}
-S^{(1)}(\phi _B) + S^{(1)}(\phi _f) + \dots) \times \nn \\
\{\frac{det'[-\nabla ^2 + U''(\phi _B)]}{det[-\nabla ^2 +
U''(\phi _f)]}\}^{-\frac{1}{2}}(1 + O[\hbar])
\label{bounce}
\eea
where the prime in the determinant denotes
omission of zero-modes, $\nabla ^2 \equiv \partial _\mu
\partial ^\mu $,
$U(\phi)$ is the field-theory potential energy,
$B=S_R(\phi _B)$,
$S_R$
is the renormalized
effective action, and $S^{(1)}$
denotes first-order renormalization
counterterms with the $\dots$ denoting higher-loop
corrections.
The bounce $\phi _B$ is
a stationary point of $S_R(\phi)$, i.e.
\be
 \partial S_R/\partial \phi _B =0
\label{station}
\ee
In our Liouville-theory analogue,
upon the identification of the renormalization
group flow with the decay of metastable vacua
in target space,
the
condition (\ref{station})
is nothing other than the
renormalization group equation (\ref{localren})
of the
$\sigma$ model action, which follows
from the renormalizability of the model
in two dimensions \cite{tseytlin}.
\pr
The ultraviolet divergence arising in the integration
over small areas in (\ref{regul}) leads to the above flow of target
time. As is shown in ref. \cite{emncpt}
this
flow is characterized by energy conservation \cite{emncpt}
and
a monotonic increase in entropy \cite{emnqm, emnuniv},
which is particularly
rapid at early times when the volume of the Universe expands
exponentially. Thus  this framework provides an {\it alternative}
to conventional field-theoretical inflation that does not
rely on an inflaton field to generate the initial
entropy of the Universe.
\pr
The flow towards the ultraviolet fixed point implies that the
apparent size of
the string world-sheet,
as measured in target space, becomes smaller as time increases.
We argue in the next section that this suppresses off-diagonal
interference terms in the configuration-space representation
of the density matrix for observable light states.

\section{Contributions to the $\nd{S}$ Matrix}
\pr
We now demonstrate that the valley configurations
discussed in section 4 make non-trivial contributions
to the $\nd{S}$ matrix, i.e. contributions that cannot be
factorized as a product of $S$ and $S^{\dagger}$ matrix elements.
We will see explicitly that these contributions diverge
logarithmically in the renormalization scale parameter, i.e.
in the time variable, as discussed in section 5. Since the
string world-sheet shrinks in target space as time progresses, these
contributions have the effect of suppressing off-diagonal elements
in the configuration-space representation of the density matrix,
removing the quantum-mechanical interference between systems
in different locations.
\pr
\nk{\it String Derivation of the $S$ matrix}
\pr
Before deriving the
contributions of the monopole-antimonopole and
instanton-anti-instanton valleys to the $\nd{S}$ matrix,
we first discuss light particle (`tachyon')
scattering in the absence of topological fluctuations
in the space-time background, showing how the $S$-matrix
of conventional quantum field theory and the
Hamiltonian evolution of conventional quantum mechanics
emerge within our treatment of target time as a
renormalization scale parameter.
\pr
We consider the
model (\ref{Yung 8}) perturbed by a `tachyon' operator
$T(X^\mu)$ : $X^\mu = (r, \theta )$ in a region of
target space far from the black hole singularity.
Since the target space is almost flat in this region,
we can Fourier transform the `tachyon' perturbation
as in equation (\ref{fourier}).
The relevant
$\sigma$-model action is
\be
S=\frac{1}{4\pi \alpha '}
\int d^2z \partial^{\alpha} X^\mu \partial_{\alpha} X^\nu \eta _{\mu\nu}
+ \int d^2z \frac{1}{a^2} T(k) e^{ik^\mu X_\mu}
\label{action}
\ee
where we
have exhibited explicitly
an ultraviolet world-sheet cut-off $a$ which appears
for dimensional
reasons. We can split the field $X^\mu$ into a zero-mode
part $x^\mu$ and a quantum part $\xi ^\mu$:
\be
X(z)^\mu =x^\mu + \xi^\mu (z)
\label{split}
\ee
and the consequent integration over $X^\mu$ in the path-integral
yields energy-momentum conservation for the `tachyon'
perturbations\footnote{At this point one should note that
in subcritical string theory ($c < 1$) the ``energies'' $k^0$ are purely
imaginary (Liouville energies), and so
the zero-mode field integration does not
imply energy conservation. However, in such a case
one can still define
an energy-conserving string theory by restricting
oneself in the residues of the poles of the
respective amplitudes \cite{polyakov}. In our case,
we are effectively working with a supercritical
string ($c > 26$), where the Liouville mode has a temporal
signature in its kinetic term, and hence energy-momentum
conservation is a consequence of the zero-mode field
integration.}.
Expanding (\ref{action})
in powers of $T(k)$, one obtains
\bea
Z &\propto& \int D\xi e^{-\int d^2z \partial^{\alpha} \xi^\mu
\partial_{\alpha} \xi^\nu \eta_{\mu\nu}}
\sum _{n} \frac{1}{n !} \int d^2z_1 \dots d^2z_n
(\frac{1}{a^2n}) \int\dots  \nn \\
\dots \int
\int d^Dk_1 &\dots& d^Dk_n \delta^{(D)}(\sum_{i}k_i)
e^{ik.\xi(z_1)}\dots e^{ik.\xi(z_n)} T(k_1)\dots T(k_n)
\label{fluctu}
\eea
This expression depends on the world-sheet cut-off,
because of ultraviolet infinities. We can exhibit
this by examining the first power of $T(k)$
in the expansion of (\ref{fluctu}). Denoting
$< \dots > \equiv \int D\xi e^{-\int \partial^\alpha  \xi ^\mu
\partial_\alpha \xi_\mu } (\dots ) $
and ignoring
the zero mode, we obtain
\bea
Z^{(1)}  \propto&\int&d^Dk
\frac{1}{a^2} T(k) \int d^2z <e^{ik.\xi (z)} >=
 \int d^2z \frac{1}{a^2} \int _k T(k) e^{-\frac{1}{2}k_\mu k_\nu
lim _{z \rightarrow z'}<\xi^\mu (z) \xi^\nu (z')>}= \nn \\
&=&\int d^2z \frac{1}{a^2} \int _k T(k) a^{\frac{k^2}{2}-2}
\label{reno}
\eea
where we have cut the flat-space propagator as follows :
\be
\    <\xi^\mu (z) \xi^\nu (0)>=\eta ^{\mu\nu} log| z + a|^2
\label{propag}
\ee
It is evident that the explicit $a$-dependence in (\ref{reno})
can be absorbed in a renormalized `tachyon' polarization
`tensor' \cite{klesussk}
\be
T_R (k)=a^{\frac{k^2}{2}-2} T(k)
\label{renocoupl}
\ee
at the lowest order in perturbation theory.
In this way we recover the leading anomalous dimension
term $\lambda$ in the perturbative renormalization group
approach to string theory.
The generic structure of the latter, in a theory with a scale $t=-ln a$,
and  renormalized
couplings $g^i(t) $ is \cite{klesussk},
\be
\beta^i = dg^i/dt = \lambda _i g^i + a^i_{jk} g^j g^k
+ \gamma ^i_{jkl} g^j g ^k g^l
\label{betaf}
\ee
where the couplings $g(t)$ are related to the bare
ones $g(0)$ by :
\be
g^i(t)=e^{\lambda_i t}g^i(0) +
[e^{(\lambda_j + \lambda_k)t} - e^{\lambda_i t}]
\frac{a^i_{jk} }{\lambda_j + \lambda_k - \lambda_i}
g^j(0) g^k(0) + b^i_{jkl}(t) g^j(0)g^k(0)g^l(0) + \dots
\label{relbare}
\ee
with
\bea
 b^i_{jkl}g^j(0)g^k(0)g^l(0)=e^{\lambda_i t}
(\frac{2a^i_{jm}a^m_{kl}}{\lambda_j + \lambda_m + \lambda_i} -
\gamma ^i_{jkl} ) \frac{1}{\lambda_j + \lambda_k + \lambda_l -
\lambda_i} + \nn \\
(\frac{2 a ^i_{jm}a^m_{kl}}{\lambda_k + \lambda_l
-\lambda_m} + \gamma^i_{jkl} ) e^{(\lambda_j + \lambda_k + \lambda_i) t}
\frac{1}{\lambda_j + \lambda_k + \lambda_l - \lambda_i}
\label{lambdas}
\eea
We recall that the lower indices in the
coefficients $a^i_{jk}$, $b^i_{jkl}$, ...
of the $\beta$-function (\ref{betaf})
are trivially symmetrized,
whilst symmetrizing the upper and lower
indices requires the Zamolodchikov
`metric'
in coupling constant space
\cite{zam},
\be
G_{ij} =2|z|^{4-\lambda_i - \lambda _j}
<[V_i (z)][ V_i(0)]>_g
\label{zammetr}
\ee
where the $[V_i]$ are normal-ordered
perturbations corresponding
to the couplings $g^i$,
and the vev is taken with respect to the
perturbed theory.
\pr
Due to the composite operators
existing in (\ref{zammetr}) there are extra infinities
in the limit $z \rightarrow 0$ that cannot be taken care by the
normal ordering of the operators $V_i$ \cite{mavrom}.
To see this,
consider the metric
(\ref{zammetr}) expressed in terms of bare
(in a
renormalization group sense)
quantities.
For the sake of
formal convenience, we shall work in the (abstract) Wilson
scheme where the {\it exact} $\beta$-function
is quadratic in the couplings\footnote{The $\epsilon$-poles
in the metric (\ref{zammetr})
can also be seen in the
practical
dimensional regularization scheme as discussed in
ref. \cite{mavrom}, to which we refer the interested reader
for details. Here we choose the Wilson approach, which is
a coupling constant expansion, since this is more relevant
for our purposes. }:
\be
\beta _{Wilson}^i =\lambda _ig^i + A^i_{jk}g^jg^k
\label{wilson}
\ee
To connect bare and renormalized quantities in this scheme
we use the concept of scaling fields $G^i$
\cite{wegner}
defined by
\be
d G^i/dt =
\lambda ^i G^i
\label{scalfields}
\ee
which are related to the
$g^i$ via :
\be
g^i= G^i + \frac{1}{2}B^i_{jk} G^jG^k + O(G^3)
\label{relationg}
\ee
with
\be
 B^i_{jk}=A^i_{jk} (\lambda_j + \lambda _k -\lambda  _i)^{-1}
\label{bcoeff}
\ee
For our purposes, an important relation is \cite{santos}
\be
\int d^2z [V_i] = \int d^2z ( V_i^B - B^j_{ik}g^k V_j^B + O(g^2))
\label{barevertex}
\ee
where the superscript `B' denotes bare quantities. Standard
scaling arguments can be used to express the two-point
function of two bare vertex operators as
\be
<V^B_i (r) V_j^B(0)>_0 =G_{ij} |z|^{\lambda _i + \lambda _j -4}
\label{twopoint}
\ee
where the subscript `0' in the vev denotes path-integral
average with respect to the fixed point action.
It can be shown \cite{santos} that the
following relation is true :
\be
G_{im}A^m_{jk}=-2\pi C_{ijk}
\label{threepoint}
\ee
where the
$C_{ijk}$ are the operator product expansion
structure constants,
which are totally symmetric in their indices.
These are universal in a renormalization group sense.
The covariant
$\beta$-function
\be
\beta_i \equiv G_{ij} \beta^j
\label{covbeta}
\ee
has, therefore,
coefficients that are totally symmetric in their
indices, and the renormalization group can be represented
as a gradient flow \cite{santos}. The corresponding flow
function is the effective action in target space \cite{mavmir2},
which generates `tachyon' scattering amplitudes
\footnote{For the case of two-dimensional $\sigma$-models
this has also been proven rigorously in
dimensional regularization,
using composite operator renormalization techniques
by means of a local renormalization-group scale
on the world-sheet
\cite{osborn}.}.
\pr
In conventional treatments
the metric (\ref{zammetr}) is defined at a {\it fixed} point
on the complex plane, $|z|=1$ \cite{santos}.
In our Liouville formalism, this no longer makes sense,
given that
the stereographic projection $\Delta R(z_1, z_2)$
on the complex plane of the
distance between two points on the world-sheet
surface changes as the area of the surface shrinks  along
the direction of the target time flow (which is opposite to the
renormalization group flow). Indeed $\Delta R(z_1, z_2)$
is given by
\be
\Delta R(z_1, z_2)=\frac{\pi}{2}\frac{|z_1 -z_2 |}
{(1 + |z_1|^2/4R^2 )^{\frac{1}{2}}
 (1 + |z_2|^2/4R^2 )^{\frac{1}{2}}}
\label{chordald}
\ee
Here $R$ is the
radius of the world-sheet, which is
assumed to have spherical topology. It is
related to the Liouville mode $\phi (z,{\overline z})$
as follows :
$\int _\Sigma \sqrt{{\hat g}} e^{-2\phi }=4\pi R^2 $,
where ${\hat g}$ is a fiducial metric on the world-sheet, which is
assumed to be
fixed \cite{david}. Close to the ultraviolet fixed
point $\phi \rightarrow \infty$ and hence $R, \Delta R$
in (\ref{chordald})
vanish. In that case the metric (\ref{zammetr})
can be shown \cite{santos}
to exhibit poles in the anomalous dimensions
of almost-marginal operators\footnote{Such small anomalous dimensions
$\lambda_i\simeq \epsilon  << 1$ serve as a regulator
for the theory, like an $\epsilon $ expansion. The same of course
is true in dimensional regularization \cite{mavrom} where
the r\^ole of the anomalous dimension is played by
the $\epsilon =d-2$, where $d$ is  the dimensionality
of the world-sheet. We do not have a precise
estimate of $\epsilon$ in our case, but note that
the $\epsilon$-expansion give successful
qualitative results even when $\epsilon$ is not small.}. To
see this,
we recall from ref. \cite{santos}
the expression for
$\partial _i G_{jk}|_{g^i=0, |z|}$:
\be
 \partial _i G_{jk}|_{g^i=0, |z|}= \frac{4\pi}{\epsilon}
 C_{ijk}(1-|z|^{2+\epsilon})
\label{dermetr}
\ee
This expression, when evaluated in standard treatments
at the point $|z|=1$, yields a vanishing coefficient
for the first-order term in a coupling-constant power
expansion for the metric. This corresponds to a
scheme choice, and is the analogue of
the diffeomorphism invariance in metric theories,
which allows the choice of a frame where the connenction
vanishes.  In our Liouville framework, the distance
$|z|$ is replaced essentially by the
stereographic projection (\ref{chordald}),
which is vanishing close to the ultraviolet fixed point,
thereby leading to an explicit example of a simple
$\epsilon$-pole
in the `metric' in coupling constant space.
It should be understood that these extra infinities
correspond to extra divergences when the
arguments of the vertex operators in (\ref{zammetr})
come close to each other. In physical expressions
one should take the finite parts
of the pertinent expressions involving $G_{ij}$.
This is always done
in the context of this work,
and will not be explicitly stated.

\pr
We now discuss the relevance  of the
string conformal
invariance
conditions to the construction of the
scattering amplitudes in target space for the light modes.
To this end,
we make a perturbative expansion of the solution
$g^i=g_0^i + g_1^i + g_2^i + \dots$ to the conformal
invariance condition $\beta^i=0$. Vanishing of the
$\beta$-function at leading order yields the mass-shell
condition $\lambda_i g_0^i =0$.
Vanishing at first order
yields
\be
g_1=-\frac{1}{\lambda_i}a^i_{jk}g_0^j g_0^k
\label{firstord}
\ee
which reproduces the three-tachyon amplitude,
whilst
the next-order vanishing condition reproduces the
four-point amplitude,
\bea
g_2=\frac{1}{\lambda_i} (\frac{2 a^i_{jm}a^m_{kl}}{\lambda_m}
-\gamma^i_{jkl})g_0^jg_0^kg_0^l= \nn \\
=\frac{1}{\lambda_i}(\frac{2 a^i_{jm}a^m_{kl}}{\lambda_m} +
D^i_{jkl} - \frac{2 a^i_{jm}a^m_{kl}}{\lambda_j +\lambda_m -\lambda_i})
g^j_0 g^k_0 g^l_0
\label{amplit}
\eea
where $D^i_{jkl}$ is the connected part.
This is the only part of the amplitude
that remains
on-shell, since the leading order
conditions imply $\lambda_i =0$ \cite{klesussk}.
\pr
In our approach described in section 5,
the ultraviolet cut-off $a$ introduced
above is interpreted as
the Liouville mode, which is in turn
identified with target time.
Thus the cut-off dependences
of couplings, etc, above should be regarded
as equivalent to conformal dressings
with the Liouville field, which lead in turn
to dependence on target time. Renormalizability
of the theory is reflected in the
independence of any physical
quantity such as the density
matrix $\rho$ from the global
cut-off scale, i.e. the {\it zero-mode}
of the Liouville field, i.e. the target time :
\be
\frac{d}{dt}\rho(g^i,p^i, t)=0
\label{deriv}
\ee
The total derivative (\ref{deriv}) may be split into
an explicit partial derivative term and an implicit
term that may be represented by the commutator
with the Hamiltonian,
so that
(\ref{deriv}) becomes
\be
\partial _t \rho =i [\rho, H]
\label{denevo}
\ee
In the case of an exactly marginal perturbation, the
corresponding $\beta$-function vanishes identically implying
that the partial time derivative $\partial \rho /\partial t = 0$
also, and equation (\ref{denevo})
becomes trivial\footnote{These vanishing conditions
should be understood in the weak
sense, i.e. as vevs of the corresponding quantum mechanical
operators between physical states.}.
This is the case of the Wheeler-De-Witt equation
in cosmology \cite{emnuniv} :
\be
[\rho (g^i, p^i, t), {\cal H}]=0
\label{dewitt}
\ee
where ${\cal H}$ is the Hamiltonian of the entire Universe.
However, as we have already discussed at length,
in our case the `tachyon' perturbation
(\ref{fourier}) is not exactly marginal,
and as a result equation (\ref{denevo})
is non-trivial. Recalling the usual
field-theoretical relationships between the
Hamiltonian, the $T$-matrix $T=\int H dt$ and the
$S$-matrix $S=1 + i T$ we see that
(\ref{denevo}), corresponds
to the conventional factorizable form
\be
\rho_{out}=\nd{S} \rho_{in} \qquad : \qquad \nd{S}=S S^\dagger
\label{factor}
\ee
of the $\nd{S}$ matrix.
\pr
The asymptotic analogue of
the Wheeler-De-Witt equation (\ref{dewitt}) is the statement
that the $T$ matrix of the full theory {\it vanishes}.
This is precisely what happens in the $SL(2,R)/U(1)$
coset model close to the singularity where the manifest
target-space $W$ symmetry \cite{emn1} is enhanced, and
contains a twisted $N=2$ world-sheet supersymmetry and hence
double bosonic $W$ symmetry \cite{emnww}. This enhanced
symmetry is understood from the observation that the model
is in fact a topological string theory
\cite{witt,eguchi,emnww,mukvaf,hor}. The fact that the
$T$ matrix of the full theory vanishes for symmetry reasons
does not mean that the theory is itself trivial. Correlation
functions for the light modes are non-trivial \cite{emnqm,mukvaf},
but this does not mean that scattering is described
by an $S$ matrix, as we now demonstrate.
\pr
\nk{\it Monopole-Antimonopole Valley Contributions
to the $\nd{S}$ Matrix}
\pr
The first non-factorizable contribution
to the $\nd{S}$ matrix that we present is that
associated with the $SL(2,R)/U(1)$
monopole-antimonopole valley.
In the two-dimensional $O(3)$ $\sigma$-model discussed
in section 4, the collective coordinates are well-known
and the action in the dilute gas approximation for instantons
has the approximate form
\be
S_{O(3)}^{(1)}(\Delta R, \rho_I, {\bar \rho}_{{\overline I}}) \propto
\frac{8\pi}{g^2} [1 - \frac{2 \rho {\bar \rho}}{(\Delta R)^2}
+ O(\frac{\rho ^2 {\bar \rho}^2}{(\Delta R)^4}) \dots ]
\label{othree}
\ee
for instanton sizes $\rho _I$ and ${\overline \rho}_{{\overline I}}$
much  smaller than the separation $\Delta R$. Note that the action
is finite in the infrared limit of large separations
$\Delta R \rightarrow \infty$. In the case of the monopole-antimonopole
valley (\ref{cocentric}) in the $SL(2,R)/U(1)$ model,
the corresponding collective coordinates
are the monopole charges $q_m$ and ${\overline q}_{{\overline m}}$
(\ref{conf}) and their relative separation $\Delta R$, which can
be projected on the complex plane by means of
the
stereographic projection (\ref{chordald}).
We work in
a phase
where the monopoles and antimonopoles
are confined
in
dipole pairs with $q_m=-{\overline q}_{{\overline m}}=q$.
The classical
action of such a well-separated
monopole-antimonopole pair takes
the following form in stereographic
coordinates:
\be
S_m=8\pi q^2 ln (2)\sqrt{2}e^\gamma
+ 2\pi q^2 ln\frac{2R}{\omega  } +
2\pi q^2 ln[\frac{|z_1-z_2|}{(4R^2 + |z_1|^2     )^{\frac{1}{2}}}
\frac{4}{(4R^2 + |z_2 |^2     )^{\frac{1}{2}}}]
\label{monact}
\ee
where $\gamma$ is Euler's constant, and
$\omega  $ is the angular
cut-off on the
world-sheet introduced in section 2,
whose radius $R$ is related in
our approach to the renormalization group
scale, i.e. the Liouville field, i.e. target time.
{\it Expression (\ref{monact}) substituted into the
generic formula (\ref{valleyap}) gives our
first non-trivial contributions to the
$\nd{S}$ matrix elements}. The action (\ref{monact})
contains first a finite part related to
the monopole charge $q$ (over which we must
integrate), that does not exhibit any particular
suppression factor, then a logarithmically-divergent
self-energy term, and finally a dipole interaction energy.
In the limit of finite separations $0 < |z_1 -z_2 |< \infty$
and very small world-sheets $R=O(a) \rightarrow 0$, the
action (\ref{monact}) yields
\be
S_m=2\pi q^2 ln \frac{a}{\omega} + finite~parts
\label{actuv}
\ee
Recalling the discussion in section 2, about the effective
variation of {\it either} $a$ or $\omega$, but {\it not both},
under the renormalization group evolution,
we observe that,
since $-ln \omega$ corresponds to target time in our approach,
equation (\ref{actuv}) exhibits an additional
time-dependence in the absorptive part
of the forward scattering amplitude,
beyond that included in the conventional
quantum Liouville equation (\ref{denevo})
obtained in the zero-defect sector, as discussed in the
previous subsection. {\it It therefore contributes
to the non-Hamiltonian term $\nd{\delta H}$ in
the equation (2) }\cite{ehns}. It should be
remarked that the correct time dependence implied
by (\ref{actuv}) should take into account quantum corrections to
dipole interactions, which are not exhibited in the
{\it classical} expression (\ref{actuv}). We shall do that
later on, by following the approach of \cite{ovrut} to
represent these effects as $\sigma$-model sine-Gordon
deformations.

\pr
This extra cut-off (time) dependence can be understood
from a world-sheet point of view. The dipole phase
of monopole-antimonopole
configurations can be represented in the $\sigma$-model
framework as a sine-Gordon
deformation \cite{ovrut} of the string action (\ref{Yung 8}).
For monopoles and antimonopoles to be confined into pairs,
as in the valley case, this deformation must be irrelevant,
breaking the conformal invariance of the $\sigma$-model action.
This introduces extra renormalization scale dependence into
`tachyon' correlation functions, causing the $\beta$-functions
for propagating light-particle operators to be non-zero.
However, as we saw in the first part of this section, target-space
scattering
amplitudes are defined by the vanishing of the respective
$\beta$-functions, and so cannot be defined
in the presence of monopole-antimonopole pairs.
\pr
The presence of $\nd{\delta H} \ne 0$ (\ref{actuv})
induces a monotonic increase in entropy (4), whose origin
can be understood intuitively as follows. Monopoles
correspond to black holes in Minkowski space, so the
monopole-antimonopole valley configuration corresponds
to a quantum fluctuation producing a microscopic
black hole in the space-time foam. Traditional quantum
gravity arguments suggested that information be `lost' across
the corresponding microscopic event horizon. In our
string approach, this information is passed by the $W$-symmetry
to unmeasured (by local scattering experiments)
massive modes (\ref{margintach}), and it is the truncation of the
effective theory to light modes that induces the
renormalization scale (time) dependence (\ref{actuv}).
\pr
\nk{\it Instanton-Anti-Instanton Valley Contribution
to the $\nd{S}$ Matrix}
\pr
The instanton-anti-instanton valley in the $SL(2,R)/U(1)$
model corresponds more closely to the $O(3)$ valley
\cite{BW,dorinst}. As in that case, one introduces
as collective coordinates the sizes $\rho _I$ and
${\overline \rho}_{{\overline I}}$ of the instanton
and anti-instanton, and their relative separation $\Delta R$.
The dilute gas approximation of section 2 corresponds
to the limit $\Delta R >> |\rho _I, {\overline \rho}_{{\overline I}} |$
in which the valley
action is approximated by a separated instanton and anti-instanton
pair with dipole interactions \cite{yung}.
The form of the latter can be inferred from that
of (\ref{Yung 33})
of the vertex operator describing an instanton-anti-instanton
pair, but its detailed form is not important for our purposes. In
contrast to the $O(3)$ $\sigma$-model case (\ref{othree}),
however, the action of an isolated instanton or anti-instanton
depends on the instanton size and the
ultraviolet cutoff, as we discussed in section
2 (see equation (\ref{instact})). Thus in the large-separation limit
of the dilute-gas approximation the single instanton-anti-instanton
valley action is
\be
S^{(1)}=kln(1 +| \rho|^2/a^2) + O(\rho {\bar \rho}/(\Delta R)^2)
\label{dipoleforce}
\ee
{\it Expression (\ref{dipoleforce}) substituted into the
generic formula (\ref{valleyap}) gives our second set of non-trivial
contributions to the $\nd{S}$ matrix elements}.
As in the monopole case, the expression (\ref{dipoleforce})
has extra renormalization scale (target time) dependence
beyond that induced in the quantum Liouville equation (\ref{denevo}),
associated in this case with the scale dependence
of $k$ discussed in section 2 (c.f. (\ref{scaledepd})):
\be
   k \propto e^{-2\gamma_0 ln|a/\Lambda |}
\label{scalek}
\ee
where $\gamma_0$ is the anomalous dimension of the
tachyon $\beta$-function.  {\it It therefore contributes
to the non-Hamiltonian  term $\nd{\delta H}$ in the
equation (2)}.
\pr
As in the monopole case, this extra cut-off (time)
dependence can be understood from the world-sheet point
of view as being due to the irrelevant  nature
(as renormalization group operators) of the deformations
(\ref{Yung 33}) that describe instanton excitations. We recall
that instantons induce transitions between black holes
with different values of $k$ (\ref{Yung 38})
and hence different masses (\ref{mass}).
Moreover, the resulting scattering operator may have extra
non-analytic behaviour for some values of $k$. Indeed, upon performing
the collective coordinate integral (\ref{valleyap}) over a region
of the size parameter $\rho$ in the action (\ref{dipoleforce})
where instanton effects are dominant, i.e. $\rho=O(a)$, we obtain
\be
\sigma_{total} \propto \int d(\Delta R) \frac{1}{k -1}
e^{E \Delta R -O(a^2/(\Delta R)^2)}
\label{instantonic}
\ee
The region of $k\simeq 1$, which is easily
reached
by instanton
transitions that can be interpreted as renormalization
group flow,
induces non-analytic
behaviour in the world-sheet correlation functions.
\pr
The origin of the increase in entropy (4)
associated with this contribution to $\nd{\delta H}$
can also be understood intuitively.
Instantons correspond to quantum jumps in the black hole mass,
interpreted in section 2 as back reaction in the presence
of light matter `tachyon' perturbations. Thus they increase
the `area' of the microscopic event horizon associated with
a microscopic black hole in the space-time foam, entailing the
loss of information. As in the monopole case, this is
passed by the $W$-symmetry to
unmeasured (by local scattering experiments)
massive modes (\ref{margintach}), whose truncation induces the
renormalization scale (time) dependence (\ref{dipoleforce}),
(\ref{scalek}).
\pr
\nk {\it Induced Von Neumann Collapse of the Density Matrix}
\pr
We now address the key question : what is the general
form of the final state density matrix $\rho_{out}$ ?
In our treatment
of time as a renormalization group scale represented by a Liouville
field, the answer to this question lies
at the ultraviolet fixed point of the renormalization
group flow. We consider the $\nd{S}$ matrix which relates
{\it in} and {\it out} states in target configuration space:
\be
{\hat \rho}_{out}(x,x') = \int dy dy' \nd{S}(x,x'; y, y')
\rho_{in}(y,y')
\label{scattering}
\ee
In the two-dimensional string approach we have taken,
light particles are represented by the massless `tachyon' mode,
which is to be regarded as a collective representation
of the $s$-wave matter fields in four dimensions. As discussed
in section 3, the $\nd{S}$ matrix in (\ref{scattering})
is related to a world-sheet tachyon correlation
function. The target space coordinates
$(x,x',y,y')$ are to be understood
as values of the zero-mode part of the
decomposition (\ref{split}) of the $\sigma$-model
variables $(X,X',Y,Y')$. This decomposition
results in a modification of the corresponding
vertex operators that represent various string backgrounds:
\bea
 &\int& d^2z g^I(X(\sigma))
 V_I(X(\sigma)) \rightarrow \nn \\
 &\rightarrow &\int d^2z \int d^Dx g^I(x)
 \delta^{(D)}(x-X(\sigma))V_I (X(\sigma)) \equiv
 \int d^2 z g^i V_i
\label{vertexoper}
\eea
where the summation over the index $i$ now includes integration
over the target-space coordinates $x$.
This notation should be understood in
the following discussion.
\pr
In our definition of target time, the latter
is a `non-relativistic' evolution
parameter of the system of light string modes\footnote{We
recall that Lorentz invariance can be regarded as an
{\it a posteriori} concept in critical strings,
which merits re-evaluation when one deals
with supercritical strings, as in our treatment
of topological defects on the world-sheet.}. The
evolution of the density matrix is given \cite{emnqm}
in the $\sigma$-model formalism (\ref{vertexoper}) by
\be
\partial _t \rho  =i[\rho, H] + i \beta^j G_{jk}[g^k, \rho]
\label{evoltemp}
\ee
We pointed out \cite{emnqm}
that this equation resembles closely the Drude model
of quantum friction discussed in ref. \cite{vernon,cald},
with the massive string modes playing the r\^oles
of `environmental oscillators'.
\pr
The analysis of ref. \cite{cald} led to the following
expression for the reduced density
matrix of the observable states:
\bea
\nonumber     \rho (g_I,g_F,  t  ) / \rho_S (g_I,g_F,  t  ) \simeq
e^{ - \eta  \int _0^{t} d\tau \int_{\tau-\epsilon}^{\tau + \epsilon}
d\tau '
\frac{(g(\tau) - g(\tau '))^2}{(\tau - \tau ')^2}} \simeq \\
e^{-\eta \int_0^{t} d\tau \int_{\tau ' \simeq \tau }
d\tau '
 \beta ^i G_{ij}(S_0) \beta ^j } \simeq
e^{ - Dt ({\bf g_I}  - {\bf  g_F} )^2 + \dots }
\label{asym}
\eea

\nk where $\eta$ is a small proportionality
constant
calculable within the Drude model\cite{vernon}, and
the subscript ``S'' denotes quantities evaluated
in conventional
Schr\"odinger
quantum mechanics. The constant
$D$ is
proportional to the sum of the
squares of the
effective anomalous dimensions of the renormalised
couplings $g^i$. As
in ref. \cite{vernon,cald} we interpret
(\ref{asym}) as
representing the overlap, after an elapsed time interval
$t$, between quantum systems localized at
different values ${\bf g}={\bf g_I}, {\bf g_F}$ of
the coordinates.
Equation (\ref{asym})
exhibits a
simple
quadratic dependence on the $\beta $-functions,
which is supported by the identification \cite{mavmir}
in string theory of the target-space effective action
with the Zamolodchikov $c$-function $C(\{ g\})$ \cite{zam}.
The latter evolves according to
\be
\partial _t C(\{ g \}) = \beta^i G_{ij} \beta^j
\label{zamol}
\ee
justifying in string theory the simple Drude model
formula (\ref{asym})
for the evolution of
density matrix
elements.
This is because
in the Drude model
of ref. \cite{vernon} the reduced density matrix
at time $t$ is given in a path-integral formalism
in terms of the `effective action' $\int _0^t d\tau
F(g^i(\tau))$ as:
\be
\rho (t) \propto exp[-\int _0 ^t d\tau  F(g^i(\tau))]
\label{effect}
\ee
In the Drude model, the
$g^i$ are `coordinates' of the
oscillators, whilst in the string
$\sigma$ model the
$g^i$ denote the light string
modes (target-space propagating fields).
Even in this case, however,
the concept of  a `coordinate' interpretation
of $g^i$ can
be maintained but in `coupling constant' space.
Identifying $F(g^i)$ with $C(g^i)$ \cite{mavmir} and using
(\ref{zamol}) in (\ref{effect}) we
recover (\ref{asym}).
\pr
We recall from ref. \cite{vernon}
that the reduced density matrix of an open
system with action $S_0(g)$ interacting
with the `environment' via the potential $V(g)$
satisfies the following relation :
\be
\rho (g_t, g'_t)=\int {\cal F}(g,g')exp(iS_0(g)-
iS_0(g'))\rho (g_0,g'_0) Dg(\tau)\dots dg'_0
\label{feynman}
\ee
where $Dg(\tau)$ denotes path integration, and  $dg_0$
ordinary integration.
The subscript $0$ and/or $t$
denote  quantities at initial
and subsequent time instants respectively. The
factor ${\cal F}(g_t,g'_t)$ is
\be
{\cal F}(g,g')=exp(-i(V(g)-V(g'))
\label{potvernon}
\ee
The integrand in (\ref{feynman}) can then be
expressed in terms of the effective
action $S_{eff}=S_0(g)-V(g)$
of the reduced subsystem, including the
interaction with the environment
\be
\rho (g_t, g'_t)=\int exp(iS_{eff}(g)-iS_{eff}(g'))
\rho (g_0,g'_0) Dg(\tau)\dots dg'_0
\label{reduced}
\ee
The important difference of the density matrix
of Feynman and Vernon (\ref{feynman})
from the reduced density
matrix (\ref{asym}) lies in the fact that the
former describes the overlap between two states
at different locations at a time $t$, while
the latter is actually a transition amplitude
from a state located at $g_I$ at time $0$
to  the same state located at $g_F$ at time $t$.
However, when considering the renormalization
group evolution as a real time evolution of
the string light-mode-subsystem, one may view
the two different locations $g$ and $g'$
appearing in the the density matrix (\ref{feynman}),
as obtained from one another by a coordinate
transformation in coupling constant space.
This in turn may be interpreted as a renormalization
scale (time) evolution, by solving the equation
\be
\Delta t = \int_{g}^{g'} \frac{\beta^i G_{ij} dg^j}
{\partial C(g)/\partial \tau}
\label{timerev}
\ee
where $C(g)$ is the Zamolodchikov $C$ function.
In this case $\rho (g_t, g'_t)$ becomes
\be
\rho (g_t,g'_t)
=\int
exp(-i\int _{0}^{t} d\tau
\beta ^i G_{ij} \beta ^j )\rho (g_0,g'_0) Dg(\tau) \dots
dg'_0
\label{cfunc}
\ee
which has a
similar structure
to
the Drude model integrand (\ref{asym}).
We then interpret this result as representing
the overlap between quantum systems
localized at different values of the
coordinates $g^i$. In the case of the
Wess-Zumino string theory studied here,
the coordinates are the `tachyon' fields $T(X)$.
Wave-function renormalization effects\footnote{It
should be remarked that such effects cannot be seen in the
first and second orders of the
$\alpha '$-expansion of perturbative
$\sigma$-models, but appear at third and higher orders
\cite{papado}. In our case we are effectively dealing with a
(formal) expansion in powers of the couplings, and hence all
orders in $\alpha '$ are taken into account.
Therefore such wave-function
renormalization effects are unambiguously present.}
in the $\sigma$-model
induce a `time' dependence on the target-space coordinates
$X \rightarrow X + \Delta X (g^i(t), t)$ in the form of
target-space diffeomorphisms \cite{papado}. The shifts
$\Delta X$
are
related to space-time
diffeomorphisms, thereby allowing the re-interpetation
of the points $X$ and $X+\Delta X$ as representing
a renormalization scheme change, so the above formalism applies.
Thus, we can express the differences between
$\sigma$-model couplings at two neighbouring points
in the $g^i$ coordinate space by a simple Taylor
expansion
\be
|g^i (t) - g'^{i}(t)|^2
\rightarrow (\nabla _X T(X))^2 |X(t)-X'(t)|^2 +\dots
\label{taylor}
\ee
We, therefore, expect the exponent in the suppression
factors (\ref{feynman}) $\rho_{out} (x,x')$ to
vanish as $x \rightarrow x'$.
\pr
Before presenting the valley contributions
to the Drude coefficient $D$, we first
make some general comments on its structure
in string theory. Recall the general decomposition  (\ref{beta})
${\hat \beta}^i=\epsilon g^i + \beta^i $ of the $\beta$-functions,
where $\beta^i=O(g^2)$ and $\epsilon \ne 0$ in the
presence of a cut-off. Recall also from the
discussion following
(\ref{zammetr}) that
the
Zamolodchikov metric,
has a pole contribution
\be
G_{ij}=\frac{1}{\epsilon}{\cal G}_{ij}^{(1)} + regular
\label{polemetric}
\ee
in the presence of a cut-off. Thus there
is a contribution to the exponent
$K$
of the Drude model (\ref{asym})
from the
$\epsilon$-term in ${\hat \beta}^i $ and the
pole term in $G_{ij}$ :
\be
K=\int _0^t {\hat \beta}^i G_{ij} {\hat \beta}^j d\tau \ni
2\int _0^t g^i{\cal G}_{ij}^{(1)}\beta^j d\tau + \epsilon
\int _0^t g^i{\cal G}_{ij}^{(1)}g^j d\tau
\label{dcoeff}
\ee
Now, using the relation $\beta^i =dg^i/dt -\epsilon g^i$
we find
\bea
K & \ni &  2 \int_0^t \int _0^t g^i{\cal G}_{ij}^{(1)}
\frac{dg^i}{d\tau} d\tau
-\epsilon \int _0^t g^i {\cal G}_{ij}^{(1)} g^j d\tau  =  \nn \\
=[[g^i(t)]^2 & - & [g^i(0)]^2] - \epsilon \int _0^t (g^i(\tau))^2 d\tau
\label{details}
\eea
Above we work in a scheme where ${\cal G_{ij}^{(1)}}$ is
constant \cite{mavrom}\footnote{In
the Wilson scheme (\ref{dermetr}) this is at least
linear in the
renormalized couplings, in which case there are cubic (and higher
order) contributions in $K$.}.
We assume by convention that $g^i(0)=0$,
and that $g^i(t)$ remains small
and $O(\epsilon )$. We extract from (\ref{details})
the time dependence of $K$, assuming
that $g^i(t)$ approaches a fixed point which is of
$O(\epsilon)$ as $t \rightarrow \infty $.
Then
\be
{\dot K} =-\epsilon [g^i(t)]^2 +  O(\epsilon^3)
\label{averagek}
\ee
and
\be
K \simeq \epsilon [g^i(t)]^2 t
\label{averaged}
\ee
for long times $t >> \frac{1}{\epsilon}$. Notice that
the suppression factor in the exponent is
linear in both time and the anomalous
dimension coefficient $\epsilon$. This is a very important
feature of the string space-time foam, that opens
the possibility of observing these non-quantum
mechanical effects
at next-generation experiments,
for instance through induced $CPT$
violation in the light-state subsystem (e.g. neutral kaons )
\cite{emncpt}.
\pr
We now show explicitly how the monopole-antimonopole
and instanton-anti-instanton valleys
contribute to the exponential
suppression factor in (\ref{asym}). We recall
from equations (\ref{monact}),(\ref{actuv})
that the monopole-antimonopole valley
contributes to the $\nd{S}$ matrix elements
a factor
\be
\nd{S} \simeq  \int exp(-S_m +\dots )
\label{actuvtwo}
\ee
where $S_m          $ is the classical
action for a monopole-antimonopole  interacting
dipole pair (\ref{monact}),(\ref{actuv}),
\be
S_m=2\pi q^2 ln  \frac{a}{\omega  } + finite~parts
\label{actuv3}
\ee
and the
integration is
over all possible separations.
The dots in (\ref{actuvtwo}) indicate dipole interactions
at finite separations. It has been shown in \cite{ovrut}
that
world-sheet dipoles can be described
by an effective action (\ref{eff})
which contains sine-Gordon type deformations of a free
$\sigma$-model action. Integration over the valley parameters
can be converted in this case into integration
over the relevant positions of the
monopole and the antimonopole.
In addition, multi-valley configurations \cite{dorinst}
can be
taken into account in a straightforward manner
by applying the formalism of
\cite{ovrut} for the description of a dipole
gas of
monopoles. In the dilute gas approximation
one can represent the result by restricting oneself
to the first power in an expansion over the sine-Gordon
deformation. Close to the ultraviolet fixed-point
(asymptotic future), one obtains the following result for
the leading monopole-antimonopole  valley contibution to the
$\nd{S}$ matrix elements :
\be
 \nd{S} \simeq
 e^{2(\Delta _m -1)ln  \omega   + \dots }
\equiv
 e^{-2(\Delta _m -1)t  + \dots }
\label{foammon}
\ee
where $\Delta _m$ is the conformal dimension
(\ref{conf})
for the sine-Gordon deformation for monopoles,
and is
proportional
to the black-hole mass. The latter is
assumed constant in the dilute
gas approximation for monopoles.
Notice that the quantum corrections
coming from dipole-interactions in (\ref{eff})
result in an extra factor of $-1$ in the
coefficient of the leading logarithmically divergent
terms of (\ref{foammon}), as compared to
the classical result (\ref{actuv3}).
The dots $\dots$ in (\ref{foammon})
represent the finite (renormalized)
valley contributions that can be computed by evaluating
averages of the sine-Gordon vertex operator \cite{ovrut}.
Their explicit form is irrelevant for our purposes.
Notice that the relation (\ref{foammon}) describes the
leading contributions close to the ultraviolet fixed
point where $\omega   \rightarrow 0$, and that
$\Delta_m -1 $ has the
interpretation of an anomalous dimension \cite{ovrut}.
Therefore the monopole-antimonopole valley suppression factor
has exactly the form of the factor $K$ in the string version of
the Drude model
(\ref{averaged}).
Because we are in the dipole phase,
$\Delta _m > 1$ and hence the
contributions (\ref{foammon}) to
the $\nd{S}$ matrix vanish
asymptotically at the
ultraviolet fixed point.
The irrelevance of the respective coupling  constants
(in this case the fugacities of the monopoles)
was crucial to the effect.
We stress the fact that the above computation
exhibits (in a generic manner) the effects
- on the light string matter -
of the
space-time foam  that are associated with the
creation of black holes in target space.
The representation of the latter by topologically
non-trivial objects (monopoles) on the world sheet
converted this complicated problem into a
simple quantum mechanical computation on the
world-sheet.
\pr
We now discuss the
instanton valleys that
represent
back-reaction effects of the matter
on the foam, i.e. transitions among black hole
states.
In the case of the instanton-anti-instanton
valley, we recall that it contributes  to the
light matter $\nd{S}$ matrix
a factor
\be
\nd{S} \simeq \frac{1}{k-1}\times (constant)
\label{instcontr}
\ee
where $k$ is renormalization-scale- (time-)
dependent according to (\ref{scaledepd}):
\be
k \simeq (\frac{a}{\Lambda})^{-2\gamma _0}
\simeq exp(-2\gamma _0 ln (|\frac{a}{\Lambda}|)
\label{scalek2}
\ee
so that
\be
\nd{S} \simeq exp( 2\gamma _0 ln (|\frac{a}{\Lambda}|)
\simeq exp(-2\gamma_0 t + \dots )
\label{dollarinstval}
\ee
The coefficient of the exponential
suppression can be interpreted as the
anomalous dimension of the tachyon matter
operator, as is
clear from the
discussion in section 2. Notice that
it is the same
anomalous dimension of the
matter deformations, though with the opposite sign,
that appears in the leading instanton contributions
to the $\nd{S}$ matrix
of the light matter. This is consistent
with the back-reaction interpretation of the instanton
effects. In the absence of any matter deformations,
such effects go away.
As in the monopole case,
this suppression factor causes
all off-diagonal entries
in the final state density matrix
to vanish  asymptotically.
Again, this
result is a consequence of the
irrelevance of the corresponding
deformations (\ref{Yung 33}) of the
string black hole $\sigma$-model.
\pr
The time-dependences of both the expressions
(\ref{foammon},\ref{dollarinstval}) mean
that monopole-antimonopole and
instanton-anti-instanton valleys
both contribute to $\nd{\delta H}$
(2,3), as in the Drude model. The
time-dependent expressions (\ref{foammon},
\ref{dollarinstval})
were both derived on the assumption that
the relevant defect and anti-defect were
far separated, and cannot be
applied directly to estimate
$\rho _{out} (x,x') $ for $(x-x') \rightarrow 0$.
However this can be inferred from the analogous
Drude model result (\ref{taylor}). In such
a case, according to our previous discussion,
the exponent in the suppression factors
(\ref{foammon},\ref{dollarinstval})
vanishes as $x \rightarrow x'$.
\pr
Thus, both monopole-antimonopole (\ref{foammon})
and
instanton-anti-instanton (\ref{dollarinstval})
valleys
yield the following result for the
density-matrix elements for light string
out-states :
\be
\rho_{out} (x, x') ={\hat \rho} (x) \delta^{(D)}(x-x')
\label{diagonal}
\ee
As already mentioned, we believe
that such contributions to the $\nd{S}$ matrix
and to the collapse (\ref{diagonal}) are
{\it generic} for string contributions to the
space-time foam.
Such a collapse of off-diagonal elements is known as
Von Neumann collapse, and
is inherent to any transition from
pure to mixed states. It
had been suggested
previously on the basis of one particular
approach to the wormhole calculus \cite{emohn2}.
The above analysis puts the conclusion
(\ref{diagonal}) on a secure basis.
It can be understood intuitively as follows. We recall
from section 5 that, in our interpretation
of time as a renormalization scale,
the string theory evolves towards
an ultraviolet fixed point
as the time $t \rightarrow \infty$.
This means that the apparent
physical size of the world-sheet,
as measured in target space,
contracts as time progresses.
This implies that interferences
between string configurations
located at different values
of the zero-mode variable $x$ vanish
asymptotically, in agreement with the explicit
calculations presented above.
\pr
We have argued previously \cite{emohn,emnqm}
that the rate
of suppression of off-diagonal entries
in the density matrix is larger
for systems containing many light particles
(`tachyons'). In particular, we found
in the Drude analogue model that $D \propto N$,
where $N$ is the number of particles in the
in- (or out-) states.
The previous arguments apply
in particular to the specific
contributions calculated above.
Moreover, in these specific cases
we can give additional
intuitive arguments.
It is well-known from other
examples that instanton and hence valley amplitudes
are larger for systems containing many particles
\cite{thooft,khoze}. Moreover, in the shrinking
world-sheet picture discussed in the previous paragraph,
it is clear that the overlap of multi-particle systems
in different locations will decrease as a product of
single-particle overlap factors. Thus quantum-mechanical
interferences vanish more rapidly for macroscopic
bodies: Schr\"odinger's cat is either
alive or dead, but not a superposition of the two!
\pr
\section{Discussion}
\pr
The original intuition behind the postulation of the $\nd{S}$
matrix and the modified quantum Liouville equation was that
pure states should evolve into mixed states within a density
matrix formalism. It was shown in \cite{ehns} that the generic effect
of such a modification of quantum mechanics would be to suppress
off-diagonal entries in the density matrix. In a single two-state
system, the general form of $\nd{\delta H}$ consistent with the
conservation of probability and energy is
\be
-\nd{\delta H} = \left( \begin{array}{c}
 \alpha \qquad  \qquad \beta \\
\beta \qquad  \qquad \gamma     \end{array}\right) \qquad : \alpha,
\gamma > 0, \alpha \gamma \ge \beta ^2
\label{hmatr}
\ee
leading to the following asymptotic form for the density matrix:
\be
\rho (t)\simeq \frac{1}{2} \left( \begin{array}{c}
    1 \qquad \qquad \qquad \qquad
exp[-(\alpha + \gamma ) t/2]exp[-i \Delta E t ] \\
    exp[-(\alpha + \gamma)t/2]exp[i \Delta E t] \qquad \qquad
\qquad \qquad 1
\end{array}\right)
\label{denstmat}
\ee
Note that the off-diagonal entries in $\rho (t)$
(\ref{denstmat}) decay exponentially, so that the
state becomes completely mixed. It was suggested
in ref. \cite{emohn} that the collapse of off-diagonal
density matrix elements should be a general feature
of this type of modification of quantum mechanics, in particular
for the configuration space representation $\rho (x, x', t)$
of the density matrix:
\be
\rho _{out} (x, x')=\rho _{in} (x', x)
[1 - D t (x -x ')^2 + \dots ]
\label{moh}
\ee
and a possible way to motivate such a collapse on the basis
of one version of the wormhole calculus was exhibited.
\pr
We have previously demonstrated \cite{emnqm}
that just such a collapse
of off-diagonal entries in the density matrix occurs
in the string
modification of quantum mechanics. We
showed that the $W_{\infty}$-induced couplings of light particles
to unobserved massive solitonic string modes induced quantum
gravitational friction leading to to the collapse of the reduced
density matrix describing the light particle system.
The formalism and conclusion resembled that of the Feynman and Vernon
description of an open quantum-mechanical system coupled to simple
harmonic oscillators representing the environment\cite{vernon,cald}.
\pr However, in our approach the collapse is an
{\it intrinsic} and
{\it inevitable} manifestation of quantum gravity that occurs
due to the nature of space-time foam and {\it independently}
of any environmental conditions. In this paper we
demonstrate this collapse phenomenon from the $\nd{S}$-matrix
point of view, using topological structures on the world-sheet
(monopoles, instantons) to represent the effects of massive string modes.
\pr
For clarity and the convenience of the reader, we summarize the
relationship of our results to general discussions
\cite{bohm, penrose}
of the
``collapse of the wave function ''. We distinguish two aspects
of this collapse: one is the suppression of off-diagonal
density matrix elements:
\be
  \rho ^{i}_{j} \rightarrow diag(p_1,p_2,\dots p_N) \qquad :
\qquad \sum_{i=1}^{N}p_i =1
\label{vncol}
\ee
generally called Von Neumann collapse, and the other is the
collapse of entities along the diagonal :
\be
diag(p_1, p_2, \dots p_N) \rightarrow diag(0,0, \dots 0, 1, 0, \dots 0)
\label{dhcol}
\ee
generally called Dirac-Heisenberg collapse. When do
these occur in conventional quantum mechanics?
In the
conventional framework an isolated quantum
mechanical system develops in a unitary manner between
measurements, given by the normal Hamiltonian evolution
embodied in the Schr\" odinger equation which becomes
the $S$ matrix asymptotically. This is called the $U$ process
in ref. \cite{penrose}. A measurement involves
a reduction of the wave function to an eigenvector
of the operator measured, and is called the $R$ process
in ref. \cite{penrose}. Both the Von Neumann (\ref{vncol})
and Dirac-Heisenberg (\ref{dhcol}) collapses are supposed
in conventional quantum mechanics to occur during the
measurement or $R$ process. The decoherence of different
eigenstates, i.e. the Von Neumann collapse (\ref{vncol}),
during a measurement has indeed been demonstrated (see, e.g.,
\cite{bohm} for a review), but the Dirac-Heisenberg
collapse (\ref{dhcol}) remains a supplementary postulate
in conventional quantum mechanics.
\pr
This conventional picture has several unsatisfactory features.
First is the question whether quantum-mechanical coherence
is retained throughout the $U$ process. Is it reasonable
for an isolated and possibly macroscopic system to exhibit
quantum-mechanical interference indefinitely ? how long can
Schr\"odinger's cat remain in a superposition of alive and dead states?
Put another way: do large objects behave classically while thay are
not observed? A second question is: whence the mysterious
mechanism of state reduction during the $R$ process?
\pr
In the modification of quantum mechanics and quantum field
theory that we derive from string, Von Neumann collapse
occurs during the $U$ process, which is therefore not unitary.
In our framework, entropy {\it increases} continuously
throughout the $U$ process as the state becomes more mixed.
This collapse occurs more rapidly for more macroscopic
systems: Schr\" odinger's cat is not in a superposition
of alive and dead states, even before we observe it. Just
such a gradual collapse between measurements has been posited
\cite{penrose} on the basis of general arguments related
to quantum gravity, but without any concrete mechanism, still
less a derivation. Subsequent to our quantum-gravitational
\cite{ehns}
approach, a stochastic mechanism for suppressing
off-diagonal density matrix elements (Von Neumann collapse)
during the $U$ process was postulated phenomenologically
in ref. \cite{ghirardi}, but without any motivation from
quantum gravity and as a supplementary postulate. We believe that
string theory also has important implications for the measurement
or $R$ process, to which we shall return in a forthcoming paper
\cite{emnmeas}. For the moment, we just note that Dirac-Heisenberg
collapse (\ref{dhcol}) involves a {\it reduction} of the entropy
of the system under consideration, which can only
occur if it is coupled to an external system (observer)
which increases its entropy sufficiently for the total entropy
of the observer and the measured system to increase overall.
Thus Dirac-Heisenberg collapse can only occur during
the measurement or $R$ process, and a string theory of measurement
is the next item on our agenda \cite{emnmeas}.
\pr
\noindent {\Large {\bf Acknowledgements }} \\

Two of us (J.E., N.E.M.) would like to thank LAPP (Annecy-le-Vieux,
France) for its hospitality while part of this work
was being done. We would also like to thank Tim Hollowood
for useful discussions.
The work of D.V.N.
is partially supported by DOE grant DE-FG05-91-ER-40633.

\newpage
\pr
{\Large {\bf Figure Captions}}
\pr
\nk{\bf Figure 1}. Stereographic projection of a two-sphere
on the complex plane. The projection of the ultraviolet
cut-off of the sphere onto the plane is indicated.
\pr
\nk{\bf Figure 2}. A diagrammatic representation of eq. (\ref{scattr}),
that represents a specific absorptive part of a world-sheet
correlation function.
\pr
\nk{\bf Figure 3}. Valley trajectory : the quantity plotted
represents the corresponding energy density.
\pr
\nk{\bf Figure 4}. Kink solution of the Euler-Lagrange
equations obtained from the reduced $SL(2,R)/U(1)$ model
(\ref{tessokto}).
\pr
\nk{\bf Figure 5}. Comparison of the scattering solution
${\overline \phi}(y,\mu)= sinh^{-1}[\mu/cosh y]$ at large $y$
with the sum of an undistorted kink and an antikink
at infinity,
$G(y,\mu)=sinh^{-1}(\frac{y + u\mu}{\sqrt{1-u^2}})-
   sinh^{-1}(\frac{y - u\mu}{\sqrt{1-u^2}})$,
approaching each other with an ultra-relativistic
velocity $u \simeq 1$.
\pr
\nk{\bf Figure 6}. Plot of the
monopole-anti-monopole valley function (\ref{cocentric})
in the $z$-plane, for $\mu \equiv v -\frac{1}{v}=1 $.
\pr
\nk{\bf Figure 7}. As in fugure 6, but for the instanton-anti-instanton
valley (\ref{confval}),
for $\mu /\sqrt{1 -u^2} =1$,
obtained from (\ref{cocentric})
by a conformal transformation. For convenience
we plot the function (\ref{confval}) on
the rescaled $z^{\frac{1}{\sqrt{1-u^2}}}$-plane.
\pr
\nk{\bf Figure 8}. Contour of integration in the analytically-continued
(regularized) version (\ref{regul}) of $\Gamma (-s)$ for $ s \in Z^+$.

\newpage
\begin{thebibliography}{99}
\bibitem{hawk} S. Hawking, Comm. Math. Phys. 87 (1982), 395.
\bibitem{ehns} J. Ellis, J.S. Hagelin, D.V. Nanopoulos and
M. Srednicki, Nucl. Phys. B241 (1984), 381.
\bibitem{bek} J. Bekenstein, Phys. Rev. D12 (1975), 3077.
\bibitem{emnqm} J. Ellis, N.E. Mavromatos
and D.V. Nanopoulos, Phys. Lett. B293 (1992), 37.
\bibitem{hawk1} S. Hawking, Comm. Math. Phys. 43 (1975), 199.
\bibitem{gross} D. Gross, Nucl. Phys. B236 (1984), 349;
see however S. Hawking, Nucl. Phys. B244 (1984), 135.
\bibitem{emohn} J. Ellis, S. Mohanty and D.V. Nanopoulos,
Phys. Lett. B221 (1989), 113.
\bibitem{emohn2} J. Ellis, S. Mohanty and D.V. Nanopoulos,
Phys. Lett. B235 (1990), 305.
\bibitem{emn1} J. Ellis, N.E.  Mavromatos and
D.V. Nanopoulos, Phys. Lett. B267 (1991), 465; {\it ibid}
B272 (1991), 261.
\bibitem{emn4d} J. Ellis, N.E.  Mavromatos and
D.V. Nanopoulos, Phys. Lett. B278 (1992), 246.
B272 (1991), 261.
\bibitem{emnloop} J. Ellis, N.E. Mavromatos and D.V. Nanopoulos,
Phys. Lett. B276 (1992), 56.
\bibitem{emnsel} J. Ellis, N.E. Mavromatos and D.V. Nanopoulos,
Phys. Lett. B284 (1992), 27; {\it ibid} B284 (1992), 43.
\bibitem{emndua} J. Ellis, N.E. Mavromatos and D.V. Nanopoulos,
Phys. Lett. B289 (1992), 25; {\it ibid} B296 91992), 40.
\bibitem{emncpt} J. Ellis, N.E. Mavromatos and
D.V. Nanopoulos, Phys. Lett. B293 (1992), 142;
CERN and Texas A \& M Univ. preprint
CERN-TH. 6755/92; ACT-24/92;CTP-TAMU-83/92.
\bibitem{emnvs} J. Ellis, N.E.  Mavromatos and
D.V. Nanopoulos, to appear.
\bibitem{emnuniv} J. Ellis, N.E. Mavromatos
and D.V. Nanopoulos, to appear.
\bibitem{zam} A.B. Zamolodchikov, JETP Lett. 43 (1986), 730;
Sov. J. Nucl. Phys. 46 (1987), 1090.
\bibitem{witt} E. Witten, Phys. Rev. D44 (1991), 314.
\bibitem{ovrut} B.A. Ovrut and S. Thomas, Phys. Rev. D43 (1991),
1314.
\bibitem{yung} A.V. Yung, SISSA preprint 97/92/EP (1992).
\bibitem{mueller} A. Mueller, Phys. Rep. 73 (1981), 237.
\bibitem{yung1} I.I. Balitsky and A.V. Yung, Phys. Lett.
B168 (1986), 113;
\par A.V. Yung, Nucl. Phys. B297 (1988), 47.
\bibitem{khoze} V.V. Khoze and A. Ringwald, Nucl.
Phys. B355 (1991), 351.
\bibitem{dorinst} N. Dorey, Los Alamos National Lab. preprint,
LA-UR-92-1380 (1992), Phys. Rev. D to be published.
\bibitem{aben} I. Antoniadis, C. Bachas, J. Ellis and
D.V. Nanopoulos, Phys. Lett. B211 (1988), 393; Nucl. Phys.
B328 (1989), 117.
\bibitem{vernon} R.P. Feynman and F.L. Vernon Jr., Ann. Phys.
(NY) 94 (1963), 118.
\bibitem{cald} A.O. Caldeira and A.J. Leggett, Ann. Phys.
149 (1983), 374.
\bibitem{bohm} D. Bohm, {\it Quantum Theory}, (Prentice-Hall,
Englewood Cliffs, N.J. (1951)), Ch. 22.
\bibitem{penrose} R. Penrose, {\it The Emperor's New Mind}
(Oxford Univ. Press, 1989).
\bibitem{abram} {\it Handbook of
Mathematical Functions}, M. Abramowitz and  I. A. Stegun
(National Bureau of Standards, 1968).
\bibitem{chaudh} S. Chaudhuri and J. Lykken, FERMILAB preprint
FERMI-PUB-92/169-T (1992).
\bibitem{tachback} J. Polchinski, Nucl. Phys. B362 (1991), 125.
\bibitem{alwis} S.P. de Alwis and J. Lykken, Phys. Lett.
B269 (1991), 264.
\bibitem{santos} N.E. Mavromatos, J.L. Miramontes and
J.M. Sanchez de Santos, Phys. Rev. D40 (1989), 535.
\bibitem{mavmir} N. Mavromatos and J.L. Miramontes,
Phys. Lett. B212 (1988), 33; N. Mavromatos, Mod. Phys.
Lett. A3 (1988), 1079; Phys. Rev. D39 (1989), 1659;
A. Tseytlin, Phys. Lett. B194 (1987), 63; H. Osborn,
Phys. Lett. B214 (1988), 555.
\bibitem{klepol} I. Klebanov and A.M. Polyakov,
Mod. Phys. Lett. A6 (1991), 3273;
\par A.M. Polyakov, Princeton Univ. preprint
PUPT-1289 (1991).
\bibitem{BW} J. Bossart and Ch. Wiesendanger,
ETH and Univ. of Z\"urich preprint, ETH-TH/91-42;
ZH-TH-32/91 (1991).
\bibitem{shore} G. Shore, Nucl. Phys. B286 (1987), 349.
\bibitem{osborn} H. Osborn in ref. \cite{mavmir}.
\bibitem{polch} J. Polchinski, Nucl. Phys. B324 (1989), 123;
\par I. Antoniadis, C. Bachas, J. Ellis and
D.V. Nanopoulos, Phys. Lett. B257 (1991), 278;
\par D.V. Nanopoulos, in {\it Proc. Int. School of
Astroparticle Physics}, HARC-Houston
(World Scientific, 1991), p. 183.
\bibitem{david} J. Distler and H. Kawai, Nucl. Phys.
B321 (1989), 509;
\par F. David, Mod. Phys. Lett. A3, (1988), 1651;
\par N.E. Mavromatos and J. L. Miramontes,
Mod. Phys. Lett. A4 (1989), 1847.
\bibitem{Moore} G. Moore and N. Seiberg,
Int. J. Mod. Phys. A7 (1991), 2601 and references therein.
\bibitem{kog} I. Kogan, Univ. of British Columbia preprint
UBCTP 91-13 (1991).
\bibitem{kutasov} M. Bershadsky and D. Kutasov, Phys. Lett.
B266 (1991), 345.
\bibitem{kogan2} I. Kogan, Phys. Lett. B265 (1991), 269.
\bibitem{bounce} S. Coleman, Phys. Rev. D15 (1977), 2929;
1248 (E) (1977).
\par C. Callan and S. Coleman, Phys. Rev. D16 91977), 1762.
\bibitem{tseytlin} A.A. Tseytlin, Nucl. Phys. B294 (1987), 383.
\bibitem{polyakov} A.M. Polyakov, Mod. Phys. Lett. A6 (1991), 635.
\bibitem{klesussk} I. Klebanov and L. Susskind,
Phys. Lett. B200 (1988), 446.
\bibitem{mavrom} N.E. Mavromatos, Phys. Rev. D39 (1989), 1659.
\bibitem{wegner} F. Wegner, in {\it Phase Transitions
and Critical Phenomena}, vol. 6,
edited M. Domb and M.S. Green
(Academic Press, New York, 1976).
\bibitem{mavmir2} N.E. Mavromatos and J.L. Miramontes,
Phys. Lett. B226 (1989), 291.
\bibitem{emnww} J. Ellis, N.E. Mavromatos and D.V. Nanopoulos,
Phys. Lett. B288 (1992), 23.
\bibitem{eguchi} T. Eguchi, Mod. Phys. Lett. A7 (1992), 85.
\bibitem{mukvaf} S. Mukhi and C. Vafa, Harvard Univ. and
Tata Inst. preprint HUTP-93/A002; TIFR/TH/93-01 (1993).
\bibitem{hor} P. Horava, Enrico Fermi Inst. preprint EFI-92-70
(1993).
\bibitem{papado} P.S. Howe, G. Papadopoulos and K.S. Stelle,
Nucl. Phys. B296 (1988), 26.
\bibitem{thooft} G. 't Hooft, Phys. Rev. Lett. 37 (1976), 8;
Phys. Rev. D14 (1976), 3432; {\it ibid} D18 (1978), 2199;
\par A. Ringwald, Nucl. Phys. B330 (1990), 1;
\par O. Espinosa, Nucl. Phys. B343 (1991), 310.
\bibitem{ghirardi} G.C. Ghirardi, A. Rimini and T. Weber,
Phys. Rev. D34 (1986), 470; G.C. Ghirardi, O. Nicrosini,
A. Rimini and T. Weber, Nuov. Cim. 102B (1988), 383.
\bibitem{emnmeas} J. Ellis, N.E. Mavromatos and D.V. Nanopoulos,
in preparation.

\end{thebibliography}
\end{document}